\documentclass[a4paper,fleqn,usenatbib]{mnras}

\usepackage{newtxtext,newtxmath}

\usepackage[T1]{fontenc}
\usepackage{ae,aecompl}

\usepackage{graphicx}	
\usepackage{amsmath}	
\usepackage{amssymb}	
\usepackage{lineno,hyperref}
\usepackage{pdflscape}

\usepackage{siunitx}
\newcommand{\percent}{\% }

\title[Pre-atmosphere meteoroid velocity accuracy]{Modeling the measurement accuracy of pre-atmosphere velocities of meteoroids}

\author[D. Vida et al.]{
Denis Vida,$^{1,2}$\thanks{E-mail: dvida@uwo.ca}
Peter G. Brown,$^{2}$
Margaret Campbell-Brown$^{2}$
\\
$^{1}$Department of Earth Sciences, University of Western Ontario, London, Ontario, N6A 5B7, Canada\\
$^{2}$Department of Physics and Astronomy, University of Western Ontario, London, Ontario, N6A 3K7, Canada\\
}

\date{Accepted 2018 July 7. Received 2018 July 5; in original form 2018 March 4}

\pubyear{2018}

\begin{document}
\label{firstpage}
\pagerange{\pageref{firstpage}--\pageref{lastpage}}
\maketitle

\begin{abstract}
Many existing optical meteor trajectory estimation methods use the approximation that the velocity of the meteor at the beginning of its luminous phase is equivalent to its velocity before atmospheric entry. Meteoroid  kinetic energy loss prior to the luminous phase cannot be measured, but for some masses and entry geometries neglecting this loss may lead to non-negligible deceleration prior to thermal ablation. Using a numerical meteoroid ablation model, we simulate the kinematics of meteoroids beginning at \SI{180}{\kilo \metre} with initial velocities ranging from $\SI{11}{\kilo \metre \per \second}$ to $\SI{71}{\kilo \metre \per \second}$, and compare model velocities at the moment of detection to measurements. We validate the simulations by comparing the simulated luminous beginning heights with observed beginning heights of different populations of meteors detected with different optical systems. We find that most low-velocity meteoroids have a significant velocity difference of \SI{100}{\metre \per \second} to \SI{750}{\metre \per \second} (depending on meteoroid type, mass, and observation system). This systematic underestimate of meteoroid speeds also results in systematically lower semi-major axes for meteoroid orbits.
\end{abstract}

\begin{keywords}
meteors -- meteoroids -- comets
\end{keywords}



\section{Introduction}

Understanding the linkage of meteor showers to their parent bodies over time requires starting conditions for backward orbital integration, namely the contemporary osculating orbits of both the parent and stream meteoroids \citep{ABEDIN2018}. However, calculation of precise heliocentric orbits of meteoroids from ground-based optical observations is difficult as atmospheric deceleration affects all measurements to some extent.  Ultimately, one has to know the pre-atmosphere position and the velocity vector of the meteoroid to a high degree of accuracy, prior to the meteoroid's interaction with the atmosphere if long-term backward integrations are to be meaningful for timescales comparable to the lifetime of a meteoroid stream. 

With the increasing precision of optical meteor observing systems, various authors have examined the probable initial velocity of meteors and performed uncertainty estimates. For example, \cite{egal2017challenge} used ``CAmera for BEtter Resolution'' (CABERNET) network data, a system which achieves a spatial precision of 3.24 arc seconds, and found that it is possible to determine meteoroid initial velocities with a precision of 1.25\percent by using the trajectory estimation method of \cite{gural2012new}. 

However, the question of true velocity accuracy is quite complex, as the velocity of the meteoroid at the beginning of its luminous phase is often equated with its pre-atmospheric velocity \citep{jenniskens2011cams, trigo20132011, segon2014draconids}, but this is not strictly true. \cite{ceplecha1987geometric} advises estimating the pre-atmosphere velocity from time vs. length along the track using the method of \cite{pecina1983new, pecina1984importance} and assumes the velocity after the correction for Earth's rotation \citep[equation 35 in][]{ceplecha1987geometric} to be equal to the no-atmospheric velocity, which may be a valid assumption for fireball-sized meteoroids, although this was never validated for fainter meteors.

For meteoroids corresponding to fireball sizes, this approach has recently been validated  by \cite{spurny2017discovery} who reduced 144 Taurid fireballs and modelled their trajectories using the \cite{ceplecha1993atmospheric} ablation model, which corresponds to the \cite{pecina1983new, pecina1984importance} model if no fragmentation is assumed. Using an atmosphere mass density model, they assumed that the velocity at the height of \SI{150}{\kilo \metre} corresponds to the pre-atmosphere velocity (private communication, Dr. Borovi\v{c}ka). They found a new branch of the Taurid meteor shower, with fireball-sized meteoroids having tightly clustered radiants/speeds as independently predicted for the Taurid resonant swarm \citep{Asher1998}. The authors quote the initial velocity of fireballs (all with initial masses higher than \SI{d-4}{\kilogram}) to within several tens of meters per second, the most precise ones approaching $\SI{\pm 7}{\metre \per \second}$. As the  \cite{ceplecha1993atmospheric} method models the full trajectory of the meteoroid, it is possible to estimate (within model assumptions) its real pre-atmosphere velocity (velocity at $t = - \infty$); indeed the authors attribute the discovery of the new Taurid branch to the high precision of their data reduction, a fact validated by the tight statistical clustering of the resonant swarm radiants. In contrast, much smaller meteoroids measured by backscatter radars \cite{brown2005velocity, brown2008meteoroid} need a deceleration correction which can be as much as \SI{6}{\kilo \metre \per \second} for meteors with beginning heights of \SI{80}{\kilo \metre}.

\cite{hajdukova2017meteoroid} have recently shown that most orbits of video meteors suffer from a significant bias in semi-major axis due to underestimated initial velocities. They point out that initial velocities of the Geminids are usually underestimated as much as \SI{200}{\metre \per \second} to \SI{500}{\metre \per \second} compared to simulations and high-precision manual reduction done by \cite{koten2004atmospheric}.

These examples motivate the general question of how accurately one can in practice measure the initial velocity of a meteoroid (i.e. the velocity at the beginning of the luminous phase) and the closely associated question of how this velocity differs from the real pre-atmosphere velocity of a meteoroid? Here we define the real pre-atmosphere velocity as the velocity prior to any sensible deceleration by the atmosphere; operationally this occurs for most meteoroids at heights above 180km.

This paper seeks to address two specific questions:
\begin{enumerate}
\item How does the true meteoroid velocity far out of the atmosphere differ from the often adopted initial velocity measured at the  beginning height as a function of mass and meteoroid type?
\item What are the effective limits to the achievable accuracy of pre-atmospheric velocities for  different optical systems and are these primarily model-related limitations or equipment limitations?
\end{enumerate}

To determine the difference between a meteoroid's velocity before it enters the atmosphere and the instrumentally measured velocity at the beginning of the luminous phase, we employ a modified single-body meteor ablation model from \cite{campbell2004model} for fainter meteors, and the FM model by \cite{ceplecha2005fragmentation} for fireballs.

In what follows, we compare simulated meteor velocities at the beginning of their modelled trajectory (at \SI{180}{\kilo \metre} height) and their velocity at the height where they would first be detected by a given optical system. We then model three real-world, but quite different, optical meteor observation systems which cover the meteoroid mass range from \SI{5d-7}{\kilogram} to $\sim$ \SI{10}{\kilogram}. The details of the modelled systems are given in section \ref{section:system_types}. 

For each system we have modelled three populations of meteoroids ranging in bulk density from \SI{180}{\kilogram \per \cubic  \metre} to \SI{5425}{\kilogram \per \cubic  \metre}. The details of the adopted material properties for each population are given in section \ref{section:meteoroid_types}. A major uncertainty in this model approach is the effect of fragmentation. Approximately $90 \%$ of faint meteors fragment during flight \citep{subasinghe2016physical}, although \cite{hawkes1975quantitative} point out that release of $\sim \SI{e-9}{\kilogram}$ grains may begin even before the luminous phase of the flight. \cite{stokan2014transverse} inspected 1800 high-resolution videos (\SI{4}{\metre} at \SI{100}{\kilo \metre} precision) recorded by the Canadian Automated Meteor Observatory of masses $\sim$ \SI{d-4}{\kilogram} \citep{weryk2013canadian} and found only 3 meteors which exhibited complex gross fragmentation which occurred before the event was recorded by the system. In what follows, we use this observation to justify use  of a single-body meteor ablation model up to the point of  detection, while using an appropriate (larger) apparent ablation coefficient to simulate continuous fragmentation into finer grains. We note that ignoring fragmentation prior to luminous onset will make our speed corrections lower limits; the true difference may be larger.

\section{Ablation models and simulation details}

\subsection{Faint meteor ablation model}

To perform our simulations for fainter meteors, we have modified the dustball model of \cite{campbell2004model} so that there is no fragmentation due to thermal disruption. 
The model assumes that the initial kinetic energy of a meteoroid is carried away by three types of energy losses: loss through heat transfer due to collisions with air molecules, black-body radiation, and heat lost with evaporating meteoroid material:

\begin{equation}
\begin{split}
    \frac{dT_m}{dt} = \frac{1}{cm} \left( \frac{\Lambda \rho_a v^3}{2} A \left( \frac{m}{\rho_m} \right)^{2/3} \right. \\
    \left. - 4 \sigma_{B} \epsilon \left( T^4_m - T^4_a \right) A  \left( \frac{m}{\rho_m} \right)^{2/3} - L \frac{dm}{dt} \right)
\end{split}
\end{equation}

\noindent where $T_m$ is the temperature of the meteoroid (initial value is assumed to be \SI{280}{\kelvin}), $c$ is the specific heat of meteoroid ($c = \SI{1000}{\joule \per \kilogram \per \kelvin}$), $m$ the meteoroid mass, $\Lambda$ the heat transfer coefficient ($\Lambda = 0.5$), $\rho_a$ the atmospheric density, which we take from the NRLMSISE-00 model (the geographical coordinates used were $\ang{45}$ N, $\ang{0}$ E on January 1, 2000 at 12:00 UTC) \citep{picone2002nrlmsise}, $v$ the meteoroid velocity, $A$ the shape factor ($A$ = 1.21, sphere), $\rho_m$ the meteoroid density, $\sigma_B$ the Stefan-Boltzmann constant, $\epsilon$ the meteoroid emissivity ($\epsilon$ = 0.9), $T_a$ the atmospheric temperature (constant at $T_a = \SI{280}{\kelvin}$) and $L$ is energy needed to ablate a unit mass (heat of ablation).

Compared to classical single-body ablation models, our model assumes that the ablation starts as the meteoroid heats high in the atmosphere, and combines the Clausius-Clapeyron partial vapour pressure equation with the additional incorporation of the Knudsen-Langmuir evaporation rate formula for calculating the mass loss:

\begin{equation}
    \frac{dm}{dt} = A \left( \frac{m}{\rho_m} \right)^{2/3} \psi \frac{P_a exp \left( \frac{L \mu}{k_B T_B} \right) exp \left( - \frac{L \mu}{k_B T_M} \right) - p_v }{ \sqrt{ \frac{2 \pi k_B T_m}{\mu} } }
\end{equation}

\noindent where $\psi$ is the condensation coefficient ($\psi$ = 0.5), $\mu$ is the molar mass ($\mu$ = \SI{36}{\atomicmassunit}), $k_B$ is the Boltzman constant, $P_a$ is the standard atmospheric pressure at sea level, $T_B$ the boiling temperature of the meteoroid material at $P_a$ ($T_B = \SI{1850}{\kelvin}$), and $p_v$ is the vapour pressure of meteoroid material at its surface (we assume $p_v = 0$ for free molecular flow, in which the meteoroids are at high altitudes).

The change in speed is calculated through conservation of momentum, when air molecules collide with the meteoroid:

\begin{equation}
    \frac{dv}{dt} = \frac{\Gamma \rho_a v^2}{m} A \left( \frac{m}{\rho_m} \right)^{2/3}
\end{equation}

\noindent here $\Gamma$ is the drag coefficient, which is assumed to be unity. Acceleration due to Earth's gravity is also taken into account.

The energy going into light production is assumed to be some fraction of the kinetic energy loss, including the deceleration term:

\begin{equation}
    I = \tau \left( \frac{dm}{dt} \frac{v^2}{2} + m v \frac{dv}{dt} \right)
\end{equation}

\noindent where $I$ is the luminous intensity and $\tau$ is the non-dimensional luminous efficiency. 

In what follows, all numerical integrations are performed using the fourth order Runge-Kutta method with a fixed time step of \SI{0.001}{\second}, until the whole mass of the meteoroid is ablated which we identify to be equivalent to the residual mass falling below $\SI{d-14}{\kilogram}$.

\subsection{Fireball ablation model} \label{subsection:fm_model}

For masses of meteoroids in the fireball range, we apply the fragmentation model (FM) by \cite{ceplecha2005fragmentation} which is based on classical single-body ablation equations with explicit addition of fragmentation. Modelling assumptions for faint meteors are not valid for larger masses, primarily because these meteoroids are no longer in free molecular flow as fireballs penetrate deeper into the atmosphere and are larger, entering the continuum flow regime \citep{campbell2004model}. The FM was developed in part to explain the discrepancies between measured photometric and dynamic masses of fireballs. In the original work it was successfully applied to 15 fireballs, and later further validated through application to meteorite-dropping fireballs \citep{borovivcka2013kovsice}.

As the FM code produces magnitudes in the photographic bandpass, we convert them to the bandpass of Sony HAD CCD based systems (see section \ref{section:allsky_system}), by applying a color index derived by \cite{silber2014observational} where $M_{HAD} = M_{ph} + 1.2$.

\section{Optical system parameters} \label{section:system_types}

To explore speed corrections using representative optical meteor observation systems in use today, we model three "typical" optical meteor systems: an image intensified system lens coupled to a CCD video camera with a narrow field of view, a moderate field of view CCD video system, and an all-sky CCD video fireball system. Each system is sensitive to a different range of meteoroid masses, peak magnitudes and beginning heights. To simulate the detectability of meteors for each system we estimate the following parameters:

\begin{itemize}
    \item The magnitude at which the system typically detects the beginning of the meteor. 
    \item The bolometric power of a zero-magnitude meteor $P_{0m}$ in each system bandpass. We assume a black-body meteor with peak temperature at $T = \SI{4500}{\kelvin}$ \citep{borovivcka2005physical} and we use Table 3 from \cite{weryk2013simultaneous} for determining $P_{0m}$ per bandpass.
    \item The typical mass of a meteoroid most commonly detected by a system is a strong function of velocity, which is determined from the observations. Mass is typically the most uncertain characteristic for a meteoroid so we appeal to the known invariance of beginning heights with meteoroid mass for smaller meteoroids \citep{hawkes1975quantitative, koten2004atmospheric}, and assume that for a range of peak magnitudes for a given optical system the corresponding meteoroid mass is purely a function of velocity.
    \item We assume a linear correlation between meteoroid velocity and peak magnitude, physical quantities which are strongly correlated (e.g. \cite{jacchia1967analysis}). Operationally, we then produce a functional fit of peak magnitude and velocity using real observations as measured by real-world examples of each type of system.
    \item For faint meteors (image intensified and moderate field of view CCD video systems) we use the \cite{campbell2004model} meteor ablation model with a fixed luminous efficiency of $\SI{0.7}{\percent}$. For fireballs (all-sky system) we use the \cite{ceplecha2005fragmentation} luminous efficiency model.
\end{itemize}

The details of the model parameters for each optical system are given in Table \ref{tab:systems_details} and described briefly in the following sections.

\begin{table*}[t]
    \centering
	\caption{Meteor limiting magnitude (MLM), equivalent bolometric power for a 0 magnitude meteor ($P_{0m}$), the expected peak magnitude (Mpeak) for meteors with a particular initial speed (V$_{init}$), estimated initial mass of a meteoroid m[kg], and assumed luminous efficiency. Models of the Canadian Automated Meteor Observatory (CAMO), Cameras for All-sky Meteor Surveillance (CAMS), and Southern Ontario Meteor Network (SOMN) are given. }
	\resizebox{\textwidth}{!}
    {
	\begin{tabular}{l c c c c c c}
	\hline\hline 
	System & Based on & MLM & $P_{0m}$ [W] & $M_{peak}$ & $\log m [kg] $ & $\tau$ \\
	\hline 
	Image intensified      & CAMO influx system, 1* & $+7.5^M$  & 840  & $-0.035 V_{init} + 4.623$ & $ \frac{ -0.4 M_{peak} }{ V^2_{init} } \log 0.098$ &  $\SI{0.7}{\%}$ \\
	Moderate field of view & CAMS, 2*              & $+5.0^M$ & 1210 & $-0.022 V_{init} + 2.244$ & 4*, modified &  $\SI{0.7}{\%}$ \\
	All-sky                & SOMN, 3*              & $-0.5^M$  & 1210 & $-0.009 V_{init} - 4.033$ & $1.8 - 3.5 \log{V_{init}} - 0.413 M_{peak}$ &  5* \\
	\hline 
	\end{tabular}
	}
	References: 1* - \cite{weryk2013canadian}, 2* - \cite{jenniskens2011cams}, 3* - \cite{brown2010development}, 4* - \cite{jacchia1967analysis}, 5* - \cite{ceplecha2005fragmentation}
	\label{tab:systems_details}
\end{table*}

\subsection{Image intensified system} \label{section:intensified_system}
The model we adopt for a narrow-field image intensified system is the Canadian Automated Meteor Observatory (CAMO) influx system as employed by the Western Meteor Physics Group (WMPG) \citep{campbell2013high, weryk2013canadian}. These systems use a high sensitivity CCD camera running at 20 frames per second with a chip of $1600 \times 1200$ pixels and 14-bit optical depth. The lens is a \SI{50}{\milli \metre} $f/0.95$ Navitar, which gives a field of view of $20 \times 20$ degrees. The camera is lens coupled to a $\SI{25}{\milli \metre}$ Generation 3 ITT model FS9925 image intensifier. The stellar limiting magnitude is $+8.5^M$, while the limiting magnitude for meteors is $+7.5^M$. 

There are two identical influx systems separated by a baseline of $\SI{45}{\kilo \metre}$, one at Elginfield ($\ang{43.193}$ N, $\ang{81.315}$ W) and the other at Tavistock ($\ang{43.264}$ N, $\ang{80.772}$ W) in Southwestern Ontario, Canada.

The photometric calibration was done in the R band for which the total bolometric power output of a zero-magnitude $T = \SI{4500}{\kelvin}$ blackbody meteor is $P_{0m} = \SI{840}{\watt}$ \citep{weryk2013simultaneous}. The magnitude and mass dependencies were fitted to 4882 manually reduced double station meteors. The trajectories were calculated using the MILIG software \citep{borovicka1990comparison} which employs the least squares line-of-sight fitting method.

The initial velocities are taken to be the average velocity of the first half of the meteor trajectory. We have only taken events with eccentricities $e < 1.0$, velocities within \SI{11.2}{\kilo \metre \per \second} and \SI{71}{\kilo \metre \per \second}, and peak magnitudes fainter than $-2^M$. 
Photometric meteoroid masses were calculated using a luminous efficiency of $\tau = 0.7 \%$ based on the integrated lightcurves. 

After performing initial meteoroid ablation simulations with these measured masses, we had to reduce the mass by a factor of 2 to match the simulation results to the observations, effectively using $\tau = 1.4 \percent$. The original photometric masses were producing events which started  at heights well above those observed. When we reduced masses further (by a factor of 3 and more), the meteors with smaller masses were too faint to be detected by the system. We attribute this to the uncertainty in the luminous efficiency $\tau$ in particular recent work which indicates that $\tau$ might be on the order of several percent for smaller meteoroids \citep{subasinghe2018luminous}. For our simulations we wish to adopt simple relations between meteoroid mass, magnitude and velocity. Hence, we performed a linear fit on the peak magnitudes versus velocities (in \SI{}{\kilo \metre \per \second}) and obtained the following relation:

\begin{equation} \label{eq:romulan_mag_fit}
    M_{peak}(V_{init}) = -0.035 V_{init} + 4.623 \pm 1.25
\end{equation}

\begin{figure}
  \includegraphics[width=\linewidth]{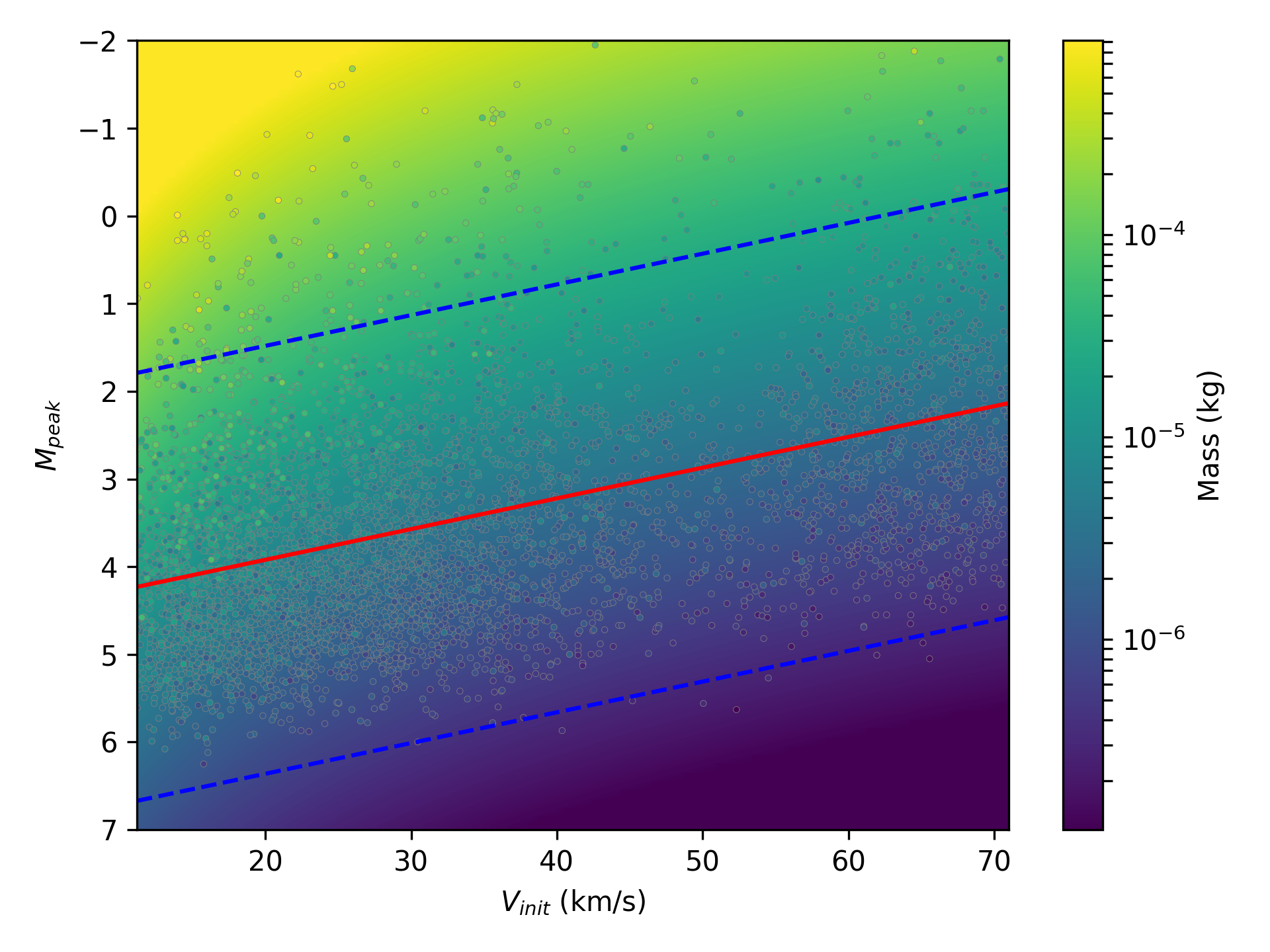}
  \caption{The photometric mass dependence as a function of the initial velocity and peak magnitude for the CAMO image intensified influx system. The colored dots represent the measurements, while the background colour represents the fit. The red line is the fit given by equation \ref{eq:romulan_mag_fit}, while blue lines represent the 95\percent confidence interval of the fit.}
  \label{fig:romulan_vel_vs_peak_mag}
\end{figure}

\begin{figure} 
  \includegraphics[width=\linewidth]{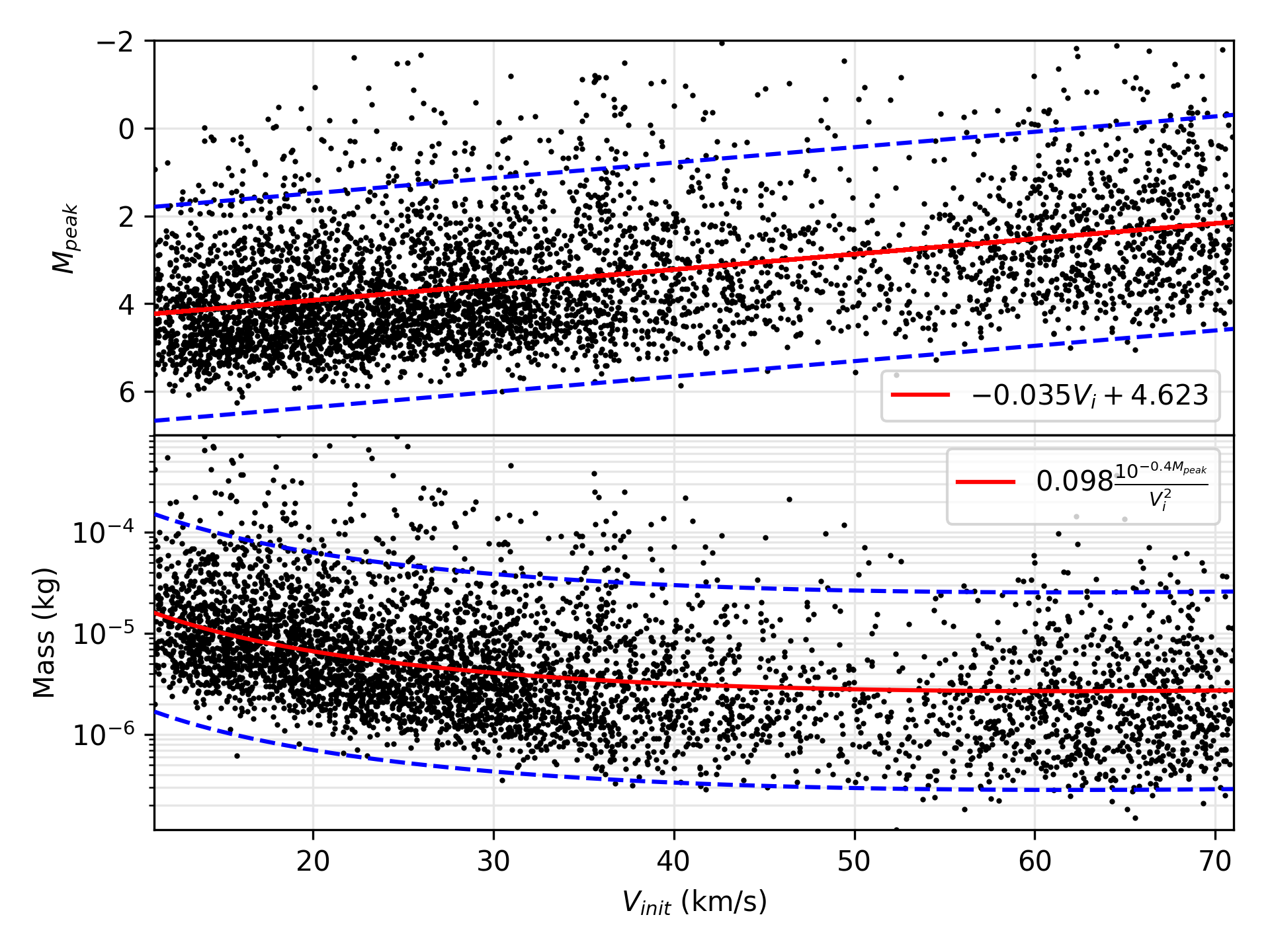}
  \caption{The top panel shows the dependence of peak magnitude on initial velocity, and the bottom plot shows the dependence of photometric mass on velocity for the CAMO image intensified influx system. Blue lines represent the 95\percent confidence interval of the fit.}
  \label{fig:romulan_masses}
\end{figure}

We also generated a simple empirical photometric mass model using this same luminous efficiency, comparing also the peak magnitudes and initial velocities:

\begin{equation}
    \log m(V_{init}, M_{peak}) = \frac{ -0.4 M_{peak} }{ V^2_{init} } \log k
\end{equation}

\noindent where photometric mass is given in kilograms and the velocity in \SI{}{\kilo \metre \per \second}. After fitting the model, we obtained $k = 0.098$. It is worth mentioning that fitting the logarithm of the photometric mass instead of the mass directly produces a better fit, as this linearizes the differences in the mass across the mass spectrum; otherwise, the fit for smaller masses is not reliable. Figure \ref{fig:romulan_vel_vs_peak_mag} shows the measured masses as colored dots, while the background color represents the model. Note that the model fits the data well, as no major discrepancies in colour between the dots and the background can be seen. Figure \ref{fig:romulan_masses} shows the dependence of the peak magnitude and the mass on the initial velocity.

\subsection{Moderate field-of-view system}

The moderate field-of-view system model is based on the Cameras for Allsky Meteor Surveillance (CAMS) \citep{jenniskens2011cams}. For this system we have used $P_{0m} = \SI{1210}{\watt}$ appropriate to a Sony HAD CCD chip \citep{weryk2013simultaneous}, used in the Watec 902 H2 Ultimate cameras operated by CAMS. Although \cite{jenniskens2011cams} state that the limiting magnitude of CAMS cameras is $+5.5^M$, they point out that very few meteors that faint are multi-station. In our simulations, we have found the value of $MLM = +5.0^M$ matches the observed beginning heights best, as we have treated the limiting magnitude as a free parameter, compensating for uncertainties in meteor geometry and the luminous efficiency.

In the CAMS orbit database \citep{jenniskens2016established}, there is no data on photometric meteoroid masses; thus we have followed the work of \cite{jenniskens2016cams} and used the results of \cite{jacchia1967analysis} to calculate meteoroid masses in grams, which we had to slightly modify as initial the simulations did not match the observations:

\begin{equation} \label{eq:jvb_mass}
\begin{split}
    \log m(V_{init}, M_{peak}, Z_G) = \log \frac{\tau_v(V_{init})}{0.03} \\
    \left(5.15 - 3.89 \log{V_{init}} - 0.33 ( M_{peak} + 0.6) - 0.67 \log{(\cos{Z_G})} \right)
\end{split}
\end{equation}

As suggested by \cite{jenniskens2016cams}, in the caption to their Table 5, we applied a color index correction of $+0.6$ to observed peak magnitudes between the photographic and HAD CCD systems before computing the mass. We also had to change the peak magnitude term from $0.44$ to $0.33$. This new value was empirically chosen because the original range of masses produced unphysical simulations - more massive meteoroids had very large beginning heights, while smaller meteoroids (fainter than peak magnitude $+3^M$) were too faint to be detected, indications that the range of masses had to be reduced. As equation \ref{eq:jvb_mass} was derived using the luminous efficiency of \cite{verniani1965luminous}, the computed masses were normalized to $\tau = 0.7\percent$, a value that produced simulations that were most consistent with observations. $\tau_v(V_{init})$ is the \cite{verniani1965luminous} luminous efficiency given as a fraction (not a percent):

\begin{equation}
    \tau_v(V_{init}) = 10^{-7} P_{v0m} V_{init}
\end{equation}

\noindent where $V_{init}$ is given in \SI{}{\kilo \metre \per \second} and $P_{v0m} = \SI{1490}{\watt}$ is the radiated power appropriate to a \SI{4500}{\kelvin} black-body zero magnitude meteor in the visual bandpass, as given by \cite{weryk2013simultaneous}.

To obtain an empirical relation between velocities and peak magnitudes for this dataset, we first filtered the CAMS data set by taking only those meteors with a convergence angle $Q_C > \ang{15}$, a reported error in geocentric velocity $\sigma_{V_g} < 10 \%$ and eccentricities $e < 1.0$. The total number of remaining meteors was 80232. A linear fit of velocity to peak magnitude produces:

\begin{equation}
    M_{peak}(V_{init}) = -0.022 V_{init} + 2.243 \pm 1.45
\end{equation}

\begin{figure} 
  \includegraphics[width=\linewidth]{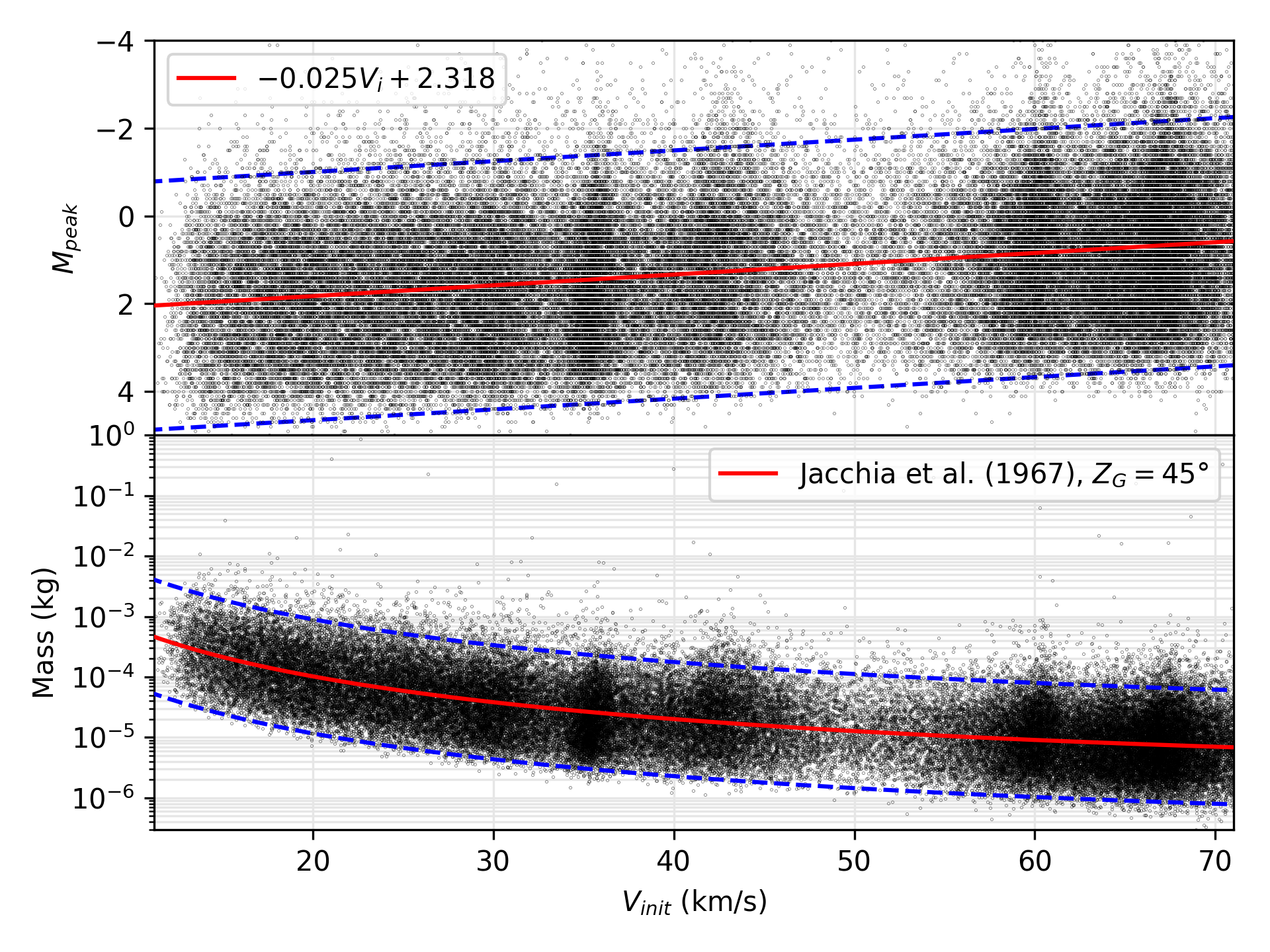}
  \caption{The top panel shows the dependence of peak magnitude on initial velocity of CAMS data, and the bottom plot shows the dependence of mass on velocity for a zenith angle $Z_G = \ang{45}$ using equation \ref{eq:jvb_mass}. Blue lines represent the 95\percent confidence interval of the fit. The horizontal banding in the top plot is due to rounding to one decimal place in the magnitude value in the original data set.}
  \label{fig:cams_masses}
\end{figure}

Figure \ref{fig:cams_masses} shows both the peak magnitude fit and the empirical mass function fit adopted in our simulations for this system.

\subsection{All-sky system} \label{section:allsky_system}

At the higher end of meteoroid masses, we investigated pre-detection decelerations of meteoroids observed by all-sky video systems. As a representative system we used the Southern Ontario Meteor Network (SOMN) \citep{weryk2008southern, brown2010development}. 

The systems use HiCam HB-710E Sony Ex-View HAD CCD cameras equipped with Rainbow L163VDC4 $1.6-3.4$\SI{}{\milli \metre} $f/1.4$ lenses. The cameras have a resolution of $640 \times 480$ pixels and are operated at 29.97 frames per second. Meteor trajectories were estimated using the method of \cite{borovicka1990comparison}. The automated data reduction pipeline only provides the average velocity of the event, though in most cases little deceleration is evident due to the low resolution of these systems. From examination of the results of the automated detection software, we find that the system most often detects meteors when they reach a visual magnitude between $0^M$ and $-1^M$. Our simulations were most consistent with observations for $MLM = -0.5^M$.

We found the automated photometry to be inconsistent with manual photometric reductions; therefore we fit our empirical relations by using representative mass and peak magnitudes from a subset of 283 manually reduced all-sky events. We have found that the peak magnitude does not show a strong correlation with velocity, probably due to saturation which occurs at higher brightness levels and the larger pixel scale of these systems:

\begin{equation} \label{eq:asgard_mag_fit}
    M_{peak}(V_{init}) = -0.009 V_{init} - 4.033 \pm 1.53
\end{equation}

\begin{figure} 
  \includegraphics[width=\linewidth]{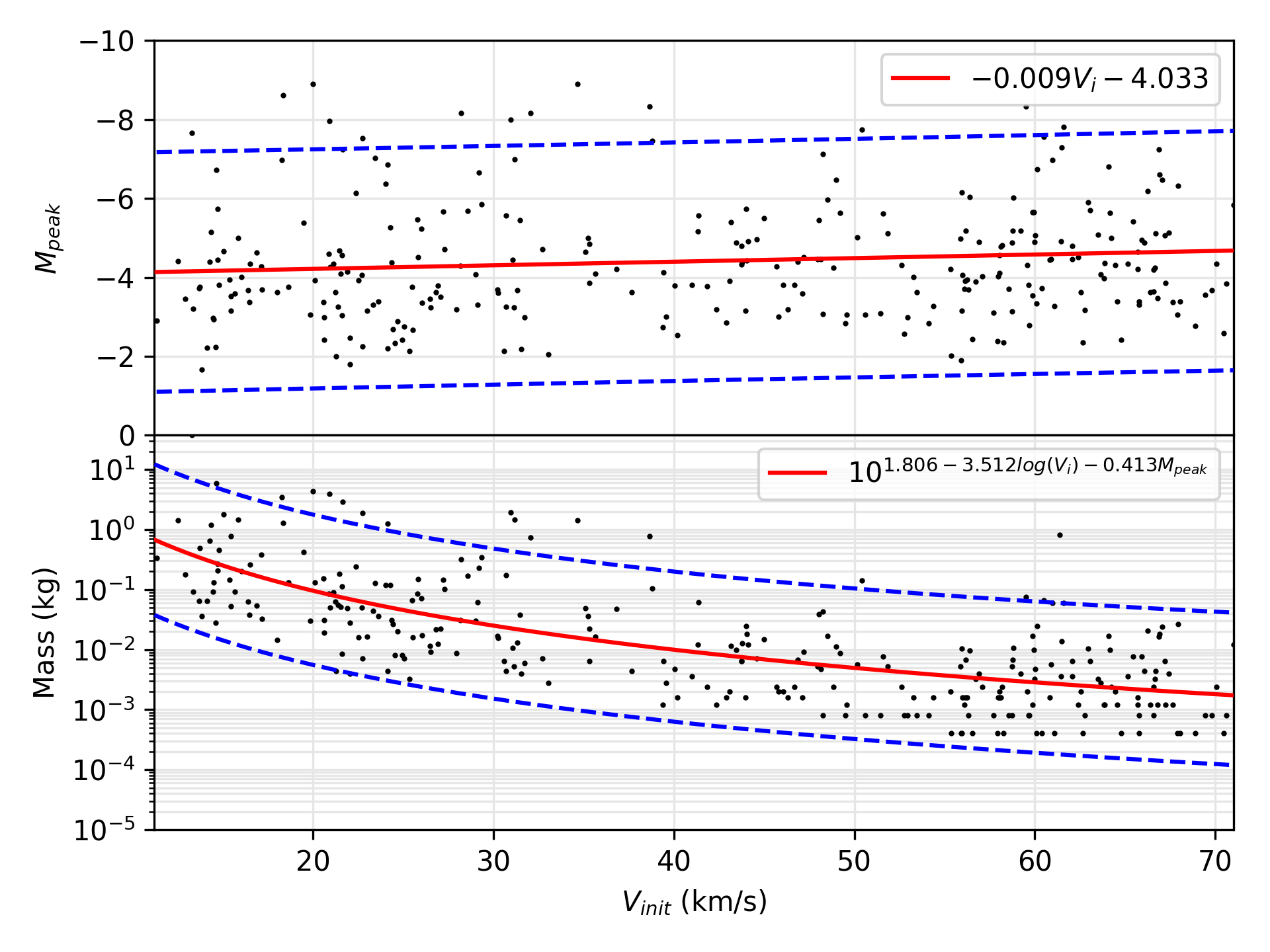}
  \caption{The top panel shows the dependence of peak magnitude on initial velocity, and the bottom plot shows the dependence of mass on velocity for the all-sky fireball system. Blue lines represent the 95\percent confidence interval of the fit. As expected for such a large pixel scale system, the average peak magnitude is a weak function of speed.}
  \label{fig:asgard_masses}
\end{figure}

In contrast to the two previous systems,  we have found that the simplistic mass model given by equation \ref{eq:romulan_mag_fit} does not fit the computed all-sky masses well, so we used a model similar to \cite{jacchia1967analysis}, but without the zenith angle term. Upon running the simulations with the original estimated photometric masses, we noticed that the smallest meteoroids are not visible to the system. Simulations matched the observations only when we increased all masses by a factor of 4, which we attribute to uncertainties in the luminous efficiency and saturation effects. The resulting mass function was the following:

\begin{equation}
    \log m(V_{init}, M_{peak}) = 1.806 - 3.512 \log{V_{init}} - 0.413 M_{peak}
\end{equation}

\noindent where the masses are given in kilograms. Figure \ref{fig:asgard_masses} shows the peak magnitude fit and the corresponding masses for values of initial velocity and peak magnitude for the subset of 283 manually measured SOMN events.

\section{Types of meteoroids} \label{section:meteoroid_types}

To cover the range of expected material properties and ablation behaviour in our model, we use three distinct types of meteoroids: cometary, asteroidal, and iron-rich. The detailed physical parameters for each category are given in Table \ref{tab:meteoroid_properties}. These classes were adopted by applying the \cite{campbell2004model} model in \cite{kikwaya2011bulk} to 107 optical observations of meteors and from model fits deriving their physical properties. Originally, \cite{kikwaya2011bulk} divided their meteoroid data into 5 types based on orbit-type, as originally proposed by \cite{borovivcka2005survey}. As our simulations are most sensitive to physical structure and not orbital information, we focus on dividing  meteoroids into density groups. 

This simple density classification scheme was motivated by figure 11 in \cite{kikwaya2011bulk} which shows a strong correlation between the meteoroid orbit Tisserand parameter with respect to Jupiter $T_J$ and meteoroid bulk density. Three distinct groupings of densities can be identified in that graph. Note that the distinction is purely by density and that meteoroids in JFC-type orbits have densities comparable to our asteroidal category, possibly indicating evolution from Asteroidal-JFC orbits through radiation forces over long timescales. We have also assumed that every meteoroid type has its own characteristic apparent ablation coefficient $\sigma$, following the classification first proposed in \cite{ceplecha1988earth}.  

Changing the apparent ablation coefficient is equivalent to adding  meteoroid fragmentation, which we have not done explicitly in the model. The apparent ablation coefficient may differ significantly from the intrinsic ablation coefficient, which does not take fragmentation into account. As shown by \cite{ceplecha2005fragmentation} the average apparent and intrinsic ablation coefficients can differ by as much as two orders of magnitude, meaning that fragmentation is the primary process of meteoroid ablation in most fireball-class (large) meteoroids. High-resolution observations of faint meteors also show a high occurrence rate of visible continuous fragmentation, indicating that the same is probably true for smaller meteoroids as well \citep{subasinghe2016physical}.

\cite{ceplecha2005fragmentation} have also shown that intrinsic ablation coefficients between different types of meteoroids are very similar, indicating that the material composition between meteoroid types is broadly similar; the ablation differences may be in bulk density and mechanical properties which only influence the rate of fragmentation \citep{borovivcka2015small}. As we use different bulk densities for the different meteoroid classes in our simulations to recreate the earliest phases of ablation, we adopt the assumptions above for the purposes of this work.

We have assumed fixed drag and heat transfer coefficients $\Gamma = 1.0$ and $\Lambda = 0.5$. The true values are uncertain and different authors have used different values: in \cite{borovivcka2007atmospheric} and \cite{fisher2000meteoroids} both values are assumed to be 1.0, while in \cite{campbell2013high} the values were $\Gamma = 1.0$ and $\Lambda = 0.4$. \cite{kikwaya2011bulk} searched values from 0.5 to 1.0 in trying to simultaneously match the dynamic and photometric measurements of their meteors. Detailed results presented in \cite{kikwaya2011phd} show no strong dependence for these values with meteoroid type. Here we use the values for drag and heat transfer given in \cite{campbell2004model}. The apparent ablation coefficient was altered only through changes to the heat of ablation $L$, thus effectively simulating different ablation rates. $L$ can be computed using the following expression:

\begin{equation}
    L = \frac{\Lambda}{2 \sigma \Gamma}
\end{equation}

The values used in our numerical entry modelling for the apparent ablation coefficients were taken from \cite{ceplecha1998meteor}, Table XVII, where meteoroid types are categorized according to \cite{ceplecha1988earth} groups: A, B, C, and D. 

Comparing that table with Table 10 in \cite{kikwaya2011bulk}, where the authors associate each Ceplecha group to their individual observed meteors, we conclude that the low-density cometary material (group C) with average density of \SI{800}{\kilogram \per \cubic \metre} has an average apparent ablation coefficient of $\sigma = \SI{0.1}{\square \second \per \square \kilo \metre}$, while the carbonaceous chondrite-like material (group A) has an ablation coefficient of $\sigma = \SI{0.042}{\square \second \per \square \kilo \metre}$. 

The properties for the iron-rich meteoroids are more uncertain; \cite{ceplecha1998meteor} gives an apparent ablation coefficient of $\sigma \approx \SI{0.07}{\square \second \per \square \kilo \metre}$ for higher densities than ours, \SI{7800}{\kilogram \per \cubic \metre}, which were derived from fireball observations in the mass range (from \SI{0.1}{\kilogram} to \SI{2d3}{\kilogram}). Due to the lack of other empirical values, we simply use $\sigma = \SI{0.07}{\square \second \per \square \kilo \metre}$ for iron-rich meteoroids, noting that for iron bodies melting as opposed to vaporiztion will dominate ablation so these larger ablation coefficients are expected. Finally, \cite{kikwaya2011bulk} find a strong correlation between the density and thermal conductivity, but because we have assumed a non-fragmenting model, thermal conductivity is not used as one of the parameters in our implementation of the \cite{campbell2004model} model.

\begin{table}
	\caption{Physical properties adopted for the three model meteoroid classes. $\rho_{min}$ and $\rho_{max}$ given the range of bulk densities of meteoroids, $\sigma$ is the apparent ablation coefficient, while $L$ is the energy needed to ablate a unit mass.}
	\resizebox{\columnwidth}{!}
    {
	\begin{tabular}{l c c c c}
	\hline\hline 
	Type & $\rho_{min}$ (\SI{}{\kilogram \per \cubic \metre}) & $\rho_{max}$ (\SI{}{\kilogram \per \cubic \metre}) & $\sigma$ (\SI{}{\square \second \per \square \kilo \metre}) & $L$ (\SI{}{\joule \per \kilogram})\\
	\hline 
	Cometary   &  180 & 1510 & 0.1   & \num{2.5d6}\\
	Asteroidal & 2000 & 3500 & 0.042 & \num{6.0d6}\\
	Iron-rich  & 4150 & 5425 & 0.07  & \num{3.6d6}\\
	\hline 
	\end{tabular}
	}
	\label{tab:meteoroid_properties}
\end{table}

\section{Simulation details} \label{section:simulation_details}

The goal of our simulation is to produce estimated brightness, speed and deceleration/mass loss profiles for a suite of meteoroids with different masses entering at a range of speeds and entry angles for all three types of meteoroids. From this simulation ``template" we then select only those meteoroids which would be detectable for a particular optical system, based on the empirical system properties summarized in Table \ref{tab:systems_details}. We then use these simulated events to compare the true initial speeds to those observed with each type of optical system.

The simulations were done in \SI{1}{\kilo \metre \per \second} steps in initial velocity $V_{\infty}$, from \SI{11}{\kilo \metre \per \second} to \SI{71}{\kilo \metre \per \second}, and across 13 zenith angle bins, from \ang{0} to \ang{75}, distributed uniformly by the cosine of the zenith angle (thus making the phase space denser at high zenith angles). For zenith angles larger than \ang{75} very few simulated meteors reached the limiting magnitude of the systems, which is consistent with observations - e.g. in CAMS data only 3\% of all orbits have zenith angles larger than \ang{75}. For very low velocity meteors (below \SI{13}{\kilo \metre \per \second}) at high zenith angles almost no ablation occurred until they were gravity accelerated to higher velocities. This often took more than \SI{10}{\second}, which we view as largely unphysical - we chose to discard these simulation runs. 

The suite of model meteor peak magnitudes were then generated by sampling in 20 uniform steps within the 95\percent confidence interval of the fit, producing 20 simulated masses. Finally, 5 uniform intervals were taken between the minimum and maximum meteoroid densities given in Table \ref{tab:meteoroid_properties} for each meteoroid type per simulated mass.

After running the meteor ablation simulation with the \cite{campbell2004model} model, luminous intensities were converted to absolute magnitudes, while the implementation of the \cite{ceplecha2005fragmentation} method provides photographic absolute magnitudes which are converted to absolute magnitudes in our bandpass (see section \ref{subsection:fm_model}). To approximate various geometries between the observers and the meteor trajectory, we have assumed that the range to the meteor at any given point corresponds to $\sqrt{(\SI{100}{\kilo \metre})^2 + h^2(t)}$, where $h(t)$ is height above the ground in kilometres. Our simulations ignore atmospheric extinction.

We define the time of the initial meteor detection $t_{init}$ as the time when the meteor's visual magnitude exceeds the system's limiting magnitude. We reject all meteors which spend less than \SI{0.15}{\second} above the detection limit. This time requirement is based on the typical value used in meteor detection algorithms, namely a meteor is detected if it is above the noise level for 4 consecutive video frames for NTSC frame rates of 30 frames per second \citep{albin2016monte}. The simulated beginning height $h_{BEG}$ of the meteor is taken as its height at time $t_{init}$. 

Similarly, the simulated measured initial velocity $v_{init}$ is the velocity at $t_{init}$. This is an upper limit to the initial velocity observed from a real optical meteor observation system, which necessarily uses a larger segment of the trail to find speeds (in most cases) during which the meteoroid will have decelerated.

For the CAMO Influx system, for example, the initial velocity is computed as the average velocity of the first half of the meteor trajectory. For the all-sky SOMN system initial velocity is equated to the average velocity across the entire trail. In both systems, these are always smaller than the real initial velocity, ie. the initial velocity at the moment the system first detects the meteor. 

For the CAMS system, the initial velocities are expected to be closer to the real initial values as they are measured with a more advanced trajectory estimation method using a global fit with time information combined with a deceleration model \citep{gural2012new}, although the real accuracy of this method remains unclear \citep{egal2017challenge}.

The difference between the starting velocity and the initial velocity is calculated for every simulation run as:

\begin{equation}
        \Delta v = v_{init} - v_{\infty} + \Delta v_{grav}
\end{equation}

\noindent where $\Delta v_{grav}$ is the change in velocity due to gravitational acceleration, which is already taken into account when computing the geocentric radiant \citep{ceplecha1987geometric}, and thus must be taken out of the total velocity difference. $\Delta v_{grav}$ was computed by running an additional no-atmosphere simulation and taking the difference between the pre-atmosphere velocity and the velocity at the height of detection.

\begin{figure}
  \includegraphics[width=\linewidth]{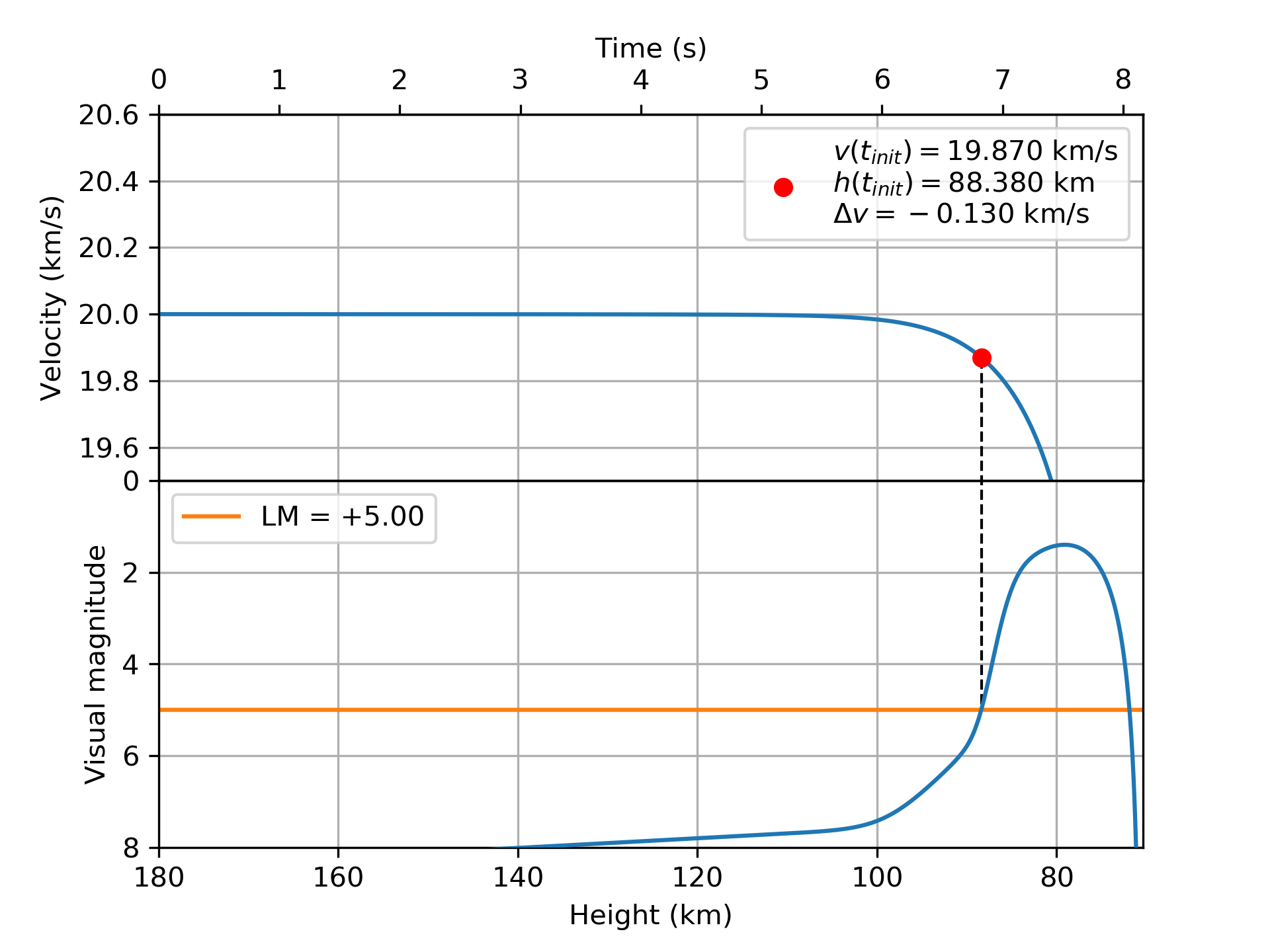}
  \caption{Ablation simulation for a $V_{\infty} = \SI{20}{\kilo \metre \per \second}$ cometary meteoroid with mass of $m = \SI{0.1}{\gram}$, density $\rho_m = \SI{1510}{\kilogram \per \cubic \metre}$ and zenith angle of $Z_G = \ang{45}$. At the limiting CAMS-like system magnitude of $MLM = +5.0^M$, the difference between the original (gravity corrected) and initial velocity was $\Delta v = \SI{-130}{\metre \per \second}$. The acceleration due to gravity was removed from the velocity in the top graph.}
  \label{fig:simulation_vel_mag}
\end{figure}

Figure \ref{fig:simulation_vel_mag} shows an example simulation for a CAMS-like system. At about $t = \SI{6}{\second}$ ablation coupled with increased atmospheric drag causes rapid deceleration. The meteor would be detected at about $t = \SI{6.7}{\second}$, when the difference from the starting velocity has reached \SI{-130}{\metre \per \second}. 

As our goal is to provide a correction for the initial speed for the entire meteoroid population for a given observation system, we averaged the velocity differences and beginning heights across simulations for all density intervals per meteoroid type. This is justified as density is not a parameter that can be easily determined from the meteor trajectory alone, as it does not correlate strongly with orbital type \citep{ceplecha1988earth}, and hence requires detailed modelling on a per event basis.

\section{Results}

To validate our working assumptions about the representative mass function and the limiting meteor magnitude for our simulated optical systems, we first compare the modelled and observed beginning heights for each optical system. We found that our results of beginnings heights versus density agree with \cite{kikwaya2011bulk}, an unsurprising result as that study used the same ablation model.

\subsection{CAMO influx system}

\begin{figure}
  \includegraphics[width=\linewidth]{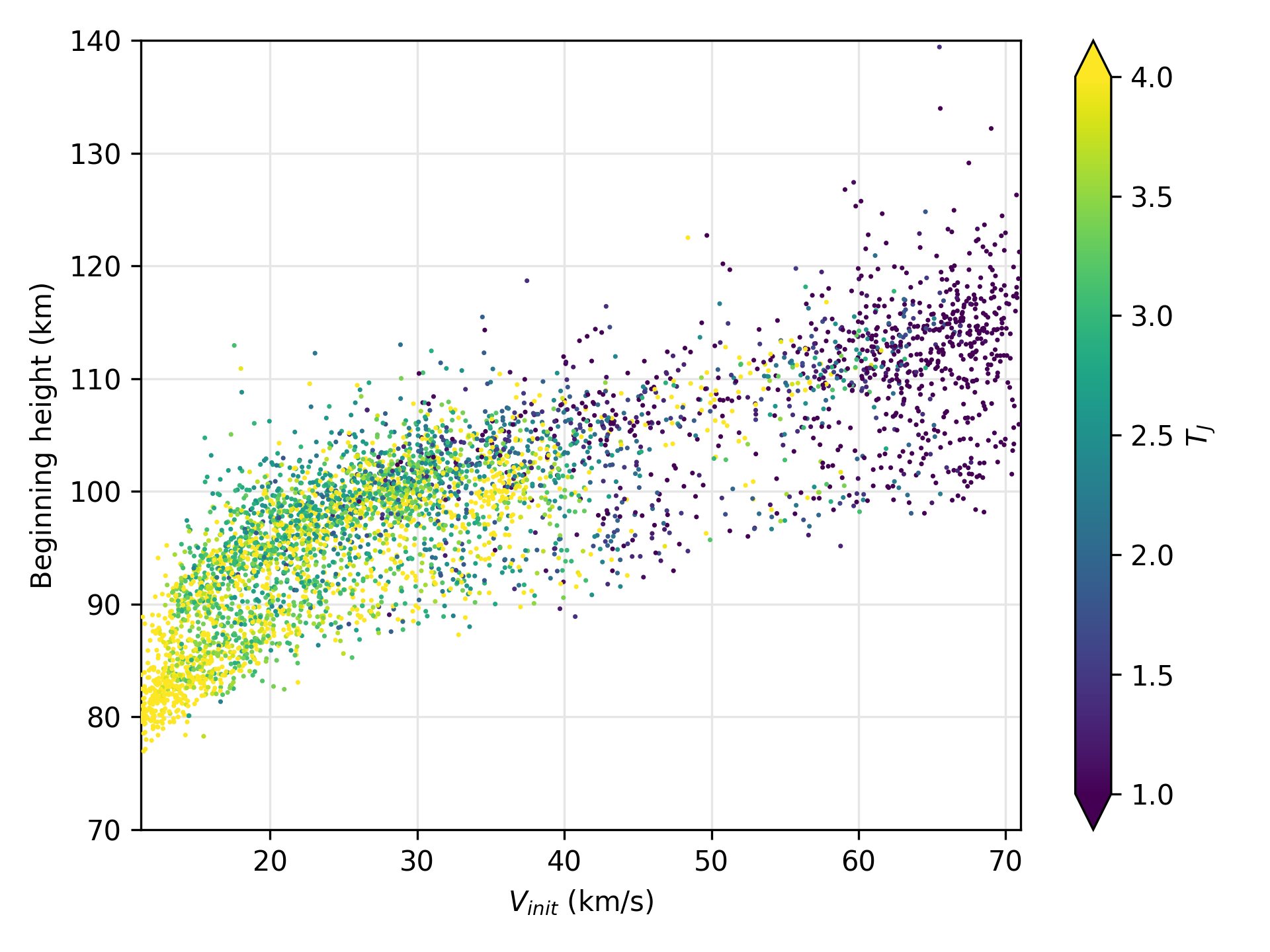}
  \caption{The observed dependence of velocity and beginning heights on the Tisserand parameter with respect to Jupiter for meteors detected by the CAMO influx system.}
  \label{fig:romulan_beg_ht}
\end{figure}

Figure \ref{fig:romulan_beg_ht} shows the observed beginning heights of real meteors imaged by the CAMO influx system as a function of speed, and their Tisserand parameters with respect to Jupiter. It can be seen that most meteoroids with  $V_{init} > \SI{40}{\kilo \metre \per \second}$ are of HTC/NIC origin, while the sub-$\SI{40}{\kilo \metre \per \second}$ ones are either JFC or asteroidal in origin. The latter dominate at the lowest velocities ($V_{init} < \SI{13}{\kilo \metre \per \second}$). This is as expected given the required orbits accessible for a given range of observed speeds at the Earth. Additionally, two branches of beginning heights can be seen, one $\sim\SI{10}{\kilo \metre}$ higher than the other \citep{ceplecha1968discrete}. Most of the observed meteors were around magnitude $+5^M$ and a large portion of them were sporadic meteors. Using the showers of the IAU Meteor Data Center for possible association, we found only 13\percent were potentially from any major shower. 

\begin{figure}
  \includegraphics[width=\linewidth]{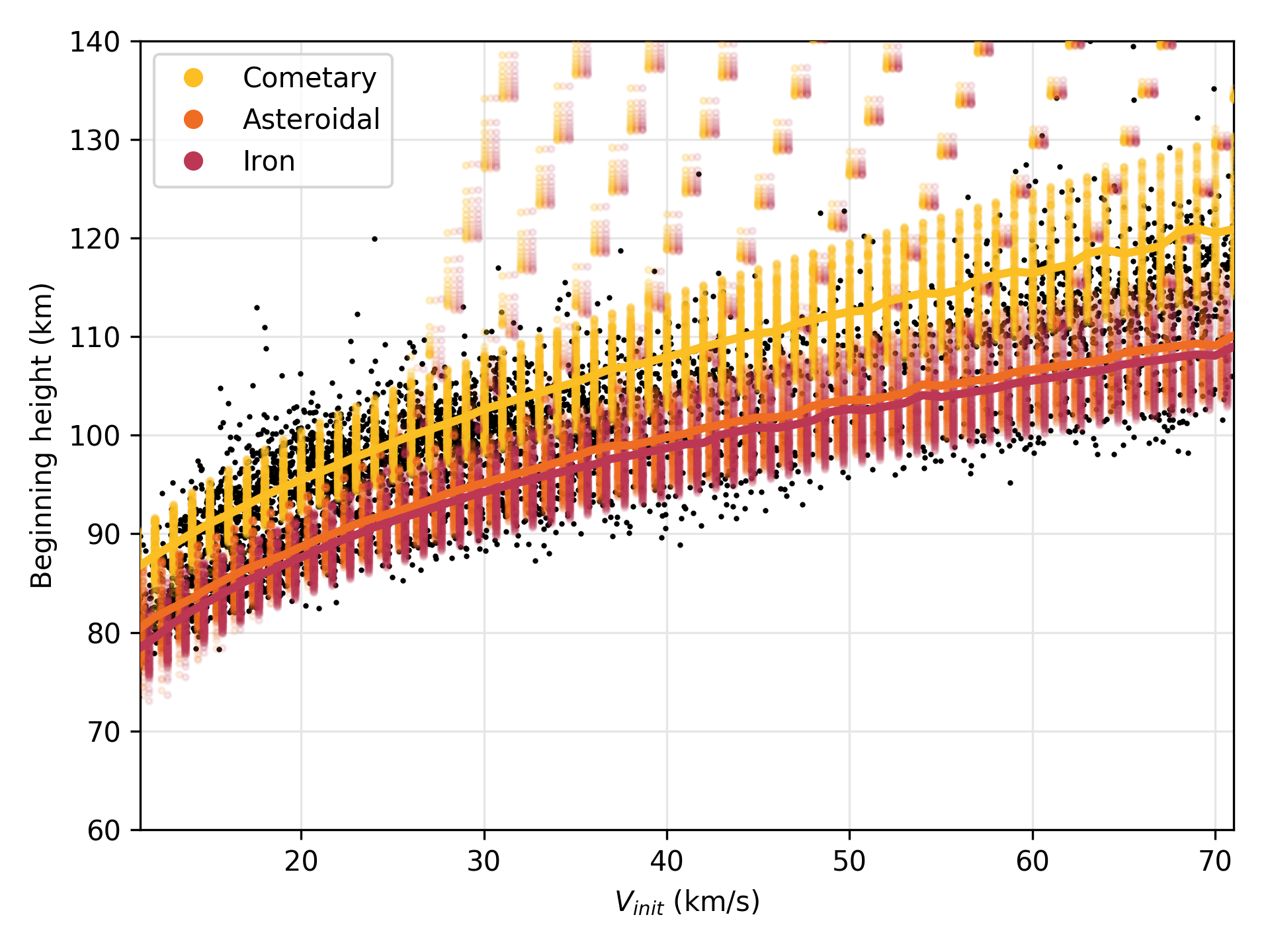}
  \caption{Comparison of the observed and simulated beginning heights for the CAMO image intensified influx system. Yellow, orange and brown dots represent cometary, asteroidal and iron meteoroids respectively. Thick lines that follow the branches by the middle are median beginning heights for every branch.}
  \label{fig:sim_intensified_beg_ht}
\end{figure}

Performing the meteor ablation simulations following the procedure described in section \ref{section:simulation_details}, the  observed and simulated beginning heights are shown in figure \ref{fig:sim_intensified_beg_ht}. The simulations generally reproduce the bulk of the observed beginning heights; lower density cometary meteoroids match the upper begin height branch, while denser asteroidal and iron meteoroids match the lower branch. These results are consistent with the findings in \cite{ceplecha1958composition} and \cite{ceplecha1968discrete}, where the higher branch was classified as type C (porous material) and the lower branch as type A (stony type). We also note that there is almost no model-predicted difference in beginning heights between asteroidal and iron-rich meteoroids. A small fraction of simulated meteors have beginning heights above the main branches. These were meteoroids with the largest masses. This is consistent with the data which show that meteors with very high beginning heights have peak magnitudes significantly brighter than the rest.

\begin{figure}
  \includegraphics[width=\linewidth]{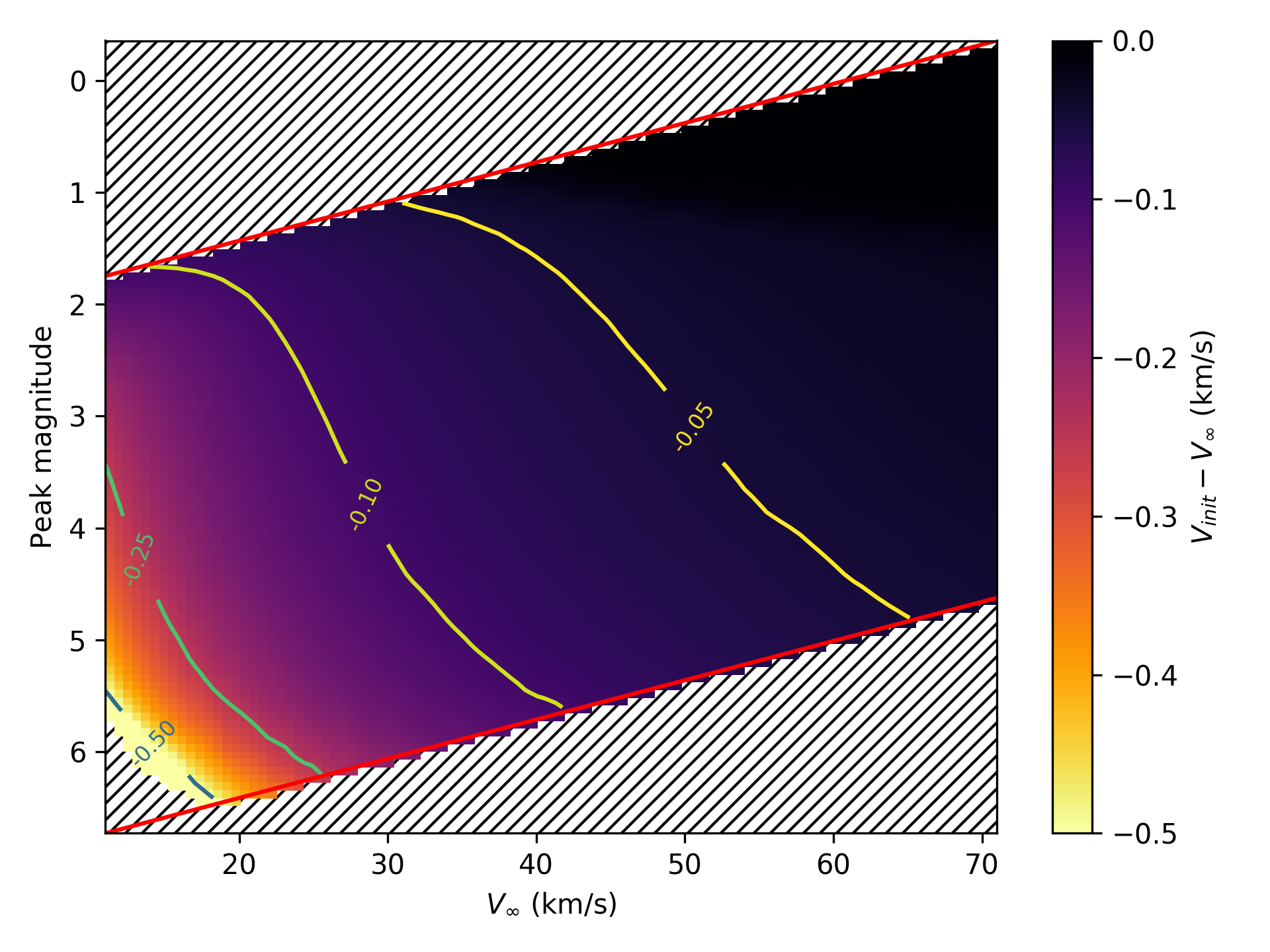}
  \caption{Cometary meteoroids - simulations for the CAMO influx image intensified systems. The areas of the parameter space which were outside of the investigated values are hatched using diagonal lines. Red lines represent the range of observed peak magnitudes, and contours ($-0.05, -0.10, -0.25, -0.5,$ and $\SI{-0.75}{\kilo \metre \per \second}$) indicate discrete values of velocity difference. The graph appears tilted due to the dependence of detectable masses on the velocity.}
  \label{fig:sim_intensified_cometary}
\end{figure}

\begin{figure}
  \includegraphics[width=\linewidth]{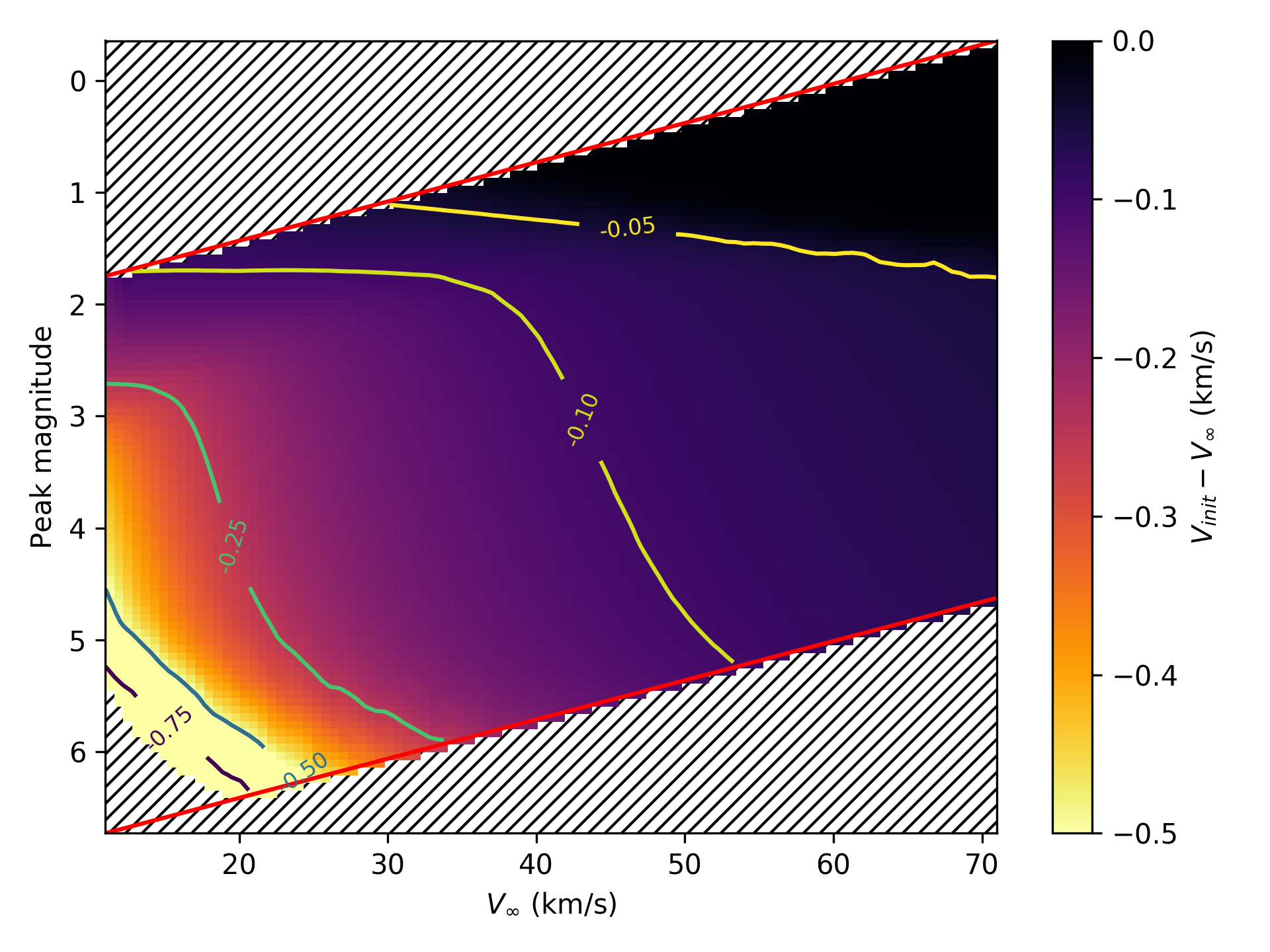}
  \caption{Asteroidal meteoroids - simulations for the image intensified systems.}
  \label{fig:sim_intensified_asteroidal}
\end{figure}

\begin{figure}
  \includegraphics[width=\linewidth]{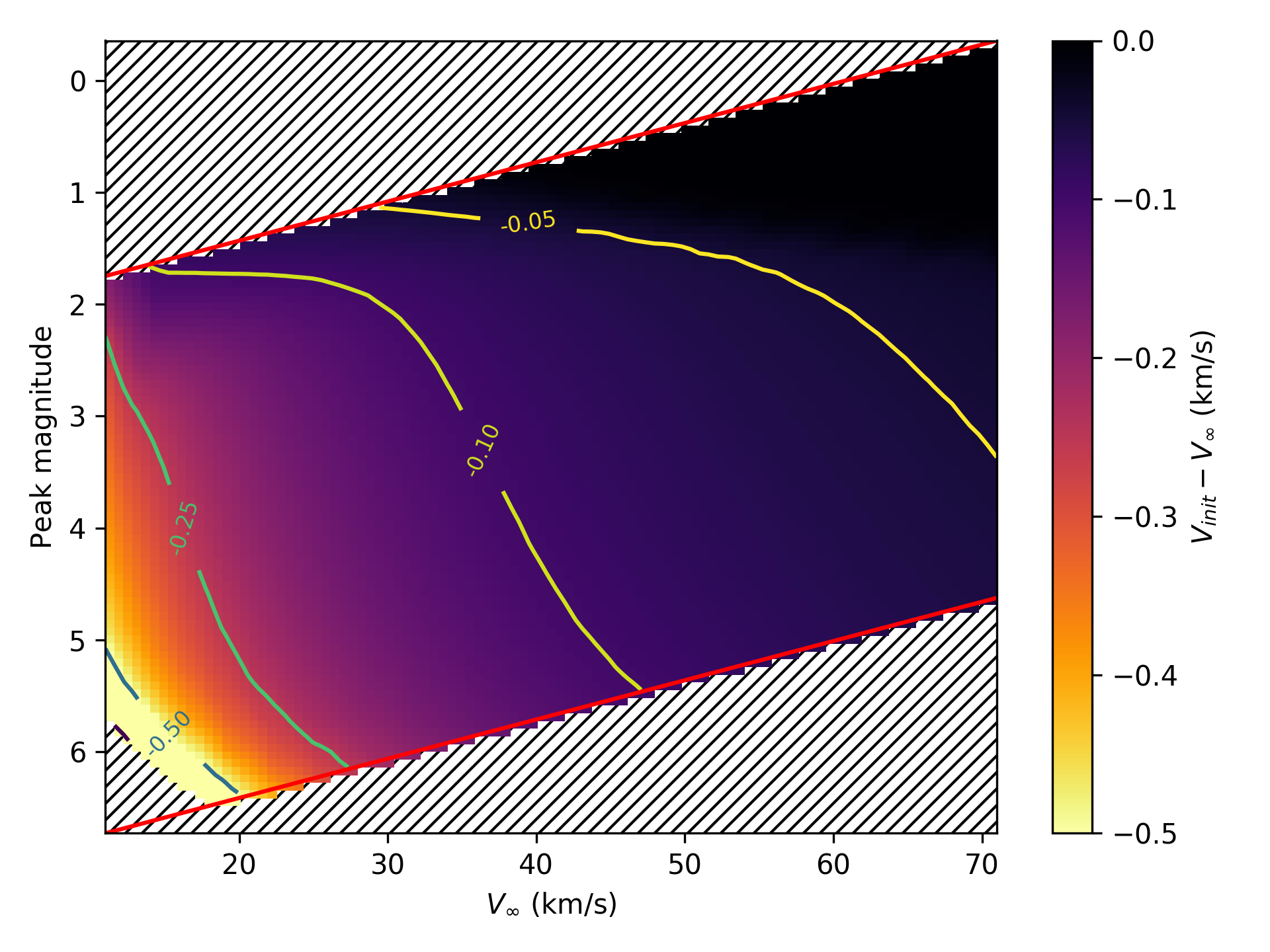}
  \caption{Iron-rich meteoroids - simulations for the CAMO influx system.}
  \label{fig:sim_intensified_iron}
\end{figure}

Figures \ref{fig:sim_intensified_cometary} through \ref{fig:sim_intensified_iron} show the simulation differences between the initial and pre-atmosphere velocities for various meteoroid types for the CAMO influx system. Overall, the deceleration at higher velocities and larger masses is only several tens of meters per second, while for lower velocities and smaller masses the velocity difference can reach several hundreds of meters per second. The influence of the zenith angle on the velocity difference is minor and not shown here, but generally $\Delta v$ increases slightly with increasing zenith angle.

There is a strong dependence of the velocity difference on the type of meteoroid material - cometary meteoroids decelerate less than asteroidal, which we believe is caused by the higher apparent ablation coefficient of cometary meteoroids and their higher beginning heights. In contrast, the velocity differences are higher for asteroidal meteoroids than for iron meteoroids, despite having similar beginning heights. We believe this is a result of higher density of iron-rich meteoroids, which leads to smaller meteoroid cross sections (where the cross section is $A \left( \frac{m}{\rho_m} \right)^{2/3}$ in the ablation equations), as we have assumed the same masses for every meteoroid type, as well as the higher apparent ablation coefficient which causes the iron-rich meteoroids to melt rather than vaporize first.




An operational fit to our results is well represented by a sixth order polynomial such that the velocity difference for any zenith angle is:

\begin{equation} \label{eq:delta_v_fit}
        \Delta v = x_0 + x_1 Z_G + x_2 Z_G^2 + x_3 Z_G^3 + x_4 Z_G^4 + x_5 Z_G^5 + x_6 Z_G^6
\end{equation}

\noindent where the zenith angle is in radians, $\Delta v$ in \SI{}{\metre \per \second}, and parameters $x_0$ to $x_6$ are given in Appendix \ref{appendix:delta_v_fits} for increments of $\SI{1}{\kilo \metre \per \second}$ in initial velocity and 20 different peak magnitudes for every meteoroid type. We note that the above relation provides the minimum correction between initial and true pre-atmospheric velocity as we have assumed no fragmentation. 

\subsection{Moderate field of view system - CAMS}

\begin{figure}
  \includegraphics[width=\linewidth]{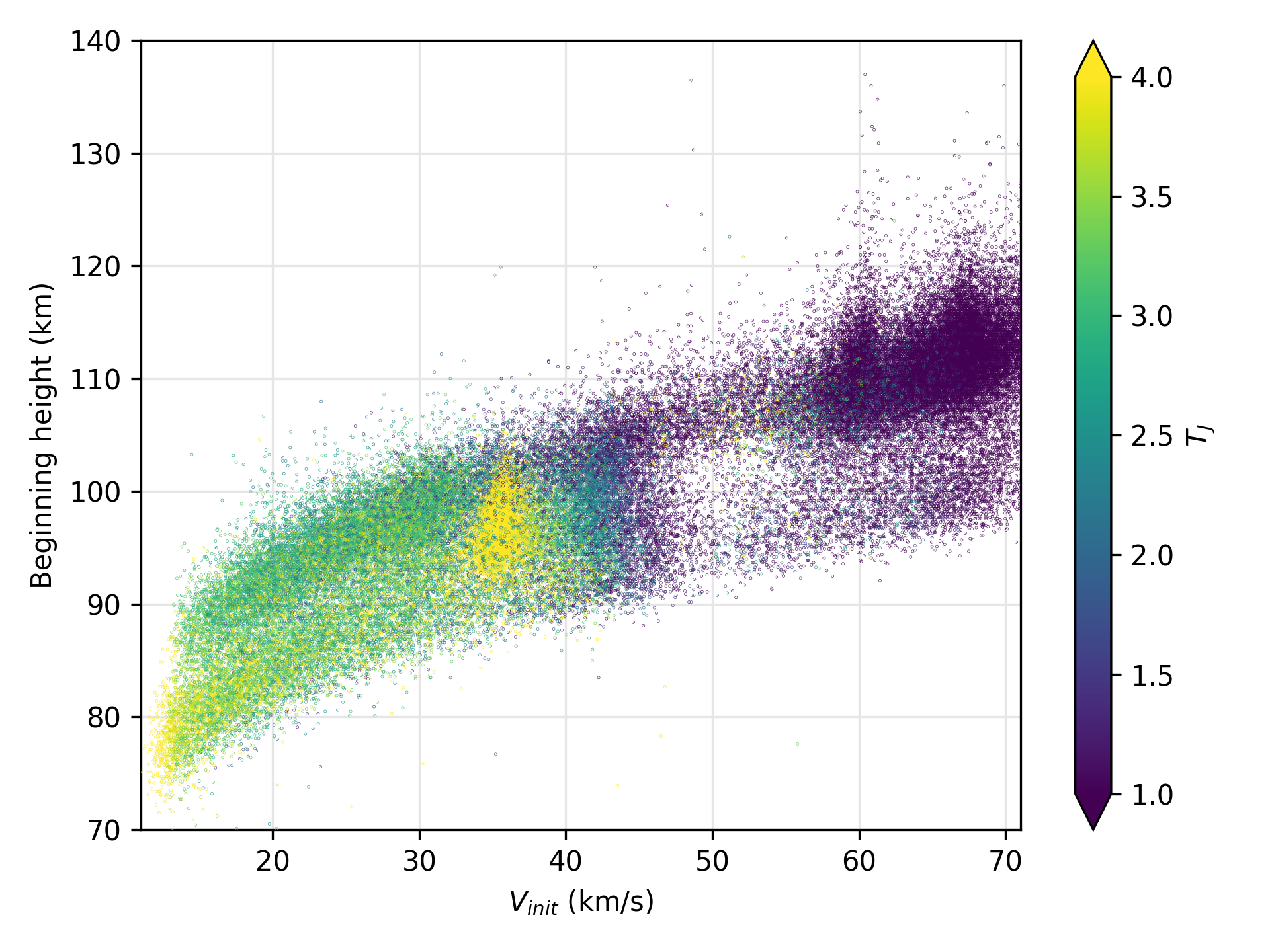}
  \caption{The dependence of velocity and beginning heights on the Tisserand parameter with respect to Jupiter for CAMS data.}
  \label{fig:cams_beg_ht}
\end{figure}

Figure \ref{fig:cams_beg_ht} shows the observed beginning heights for actual CAMS optical system meteors as a function of speed, and their Tisserand parameters with respect to Jupiter. The same distinction in orbit - types for $V_{init} < \SI{40}{\kilo \metre \per \second}$ and $V_{init} > \SI{40}{\kilo \metre \per \second}$ meteors can be seen here as was present with the CAMO influx system, as well as the same separation into two branches of beginning heights. These data contain a substantially larger fraction of shower meteors (27.6 \percent), notably the Geminids at $V_{init} \approx \SI{35}{\kilo \metre \per \second}$, and Perseids and Orionids at $V_{init} \approx \SI{60}{\kilo \metre \per \second}$ and $V_{init} \approx \SI{66}{\kilo \metre \per \second}$ respectively.

\begin{figure}
  \includegraphics[width=\linewidth]{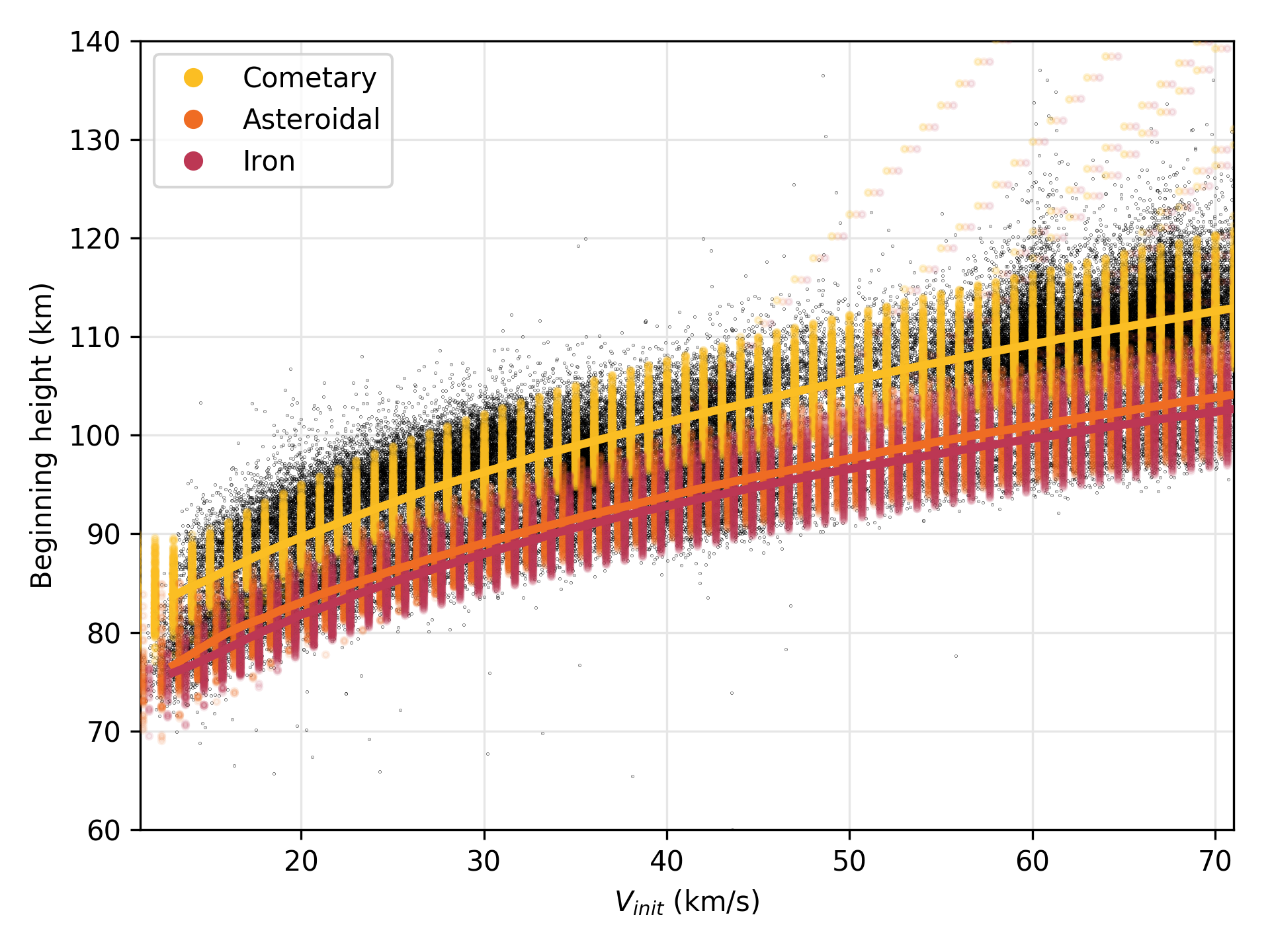}
  \caption{Observed and simulated beginning heights for the CAMS-type system. Yellow, orange and brown dots represent cometary, asteroidal and iron meteoroids respectively. Thick lines that follow the branches by the middle are median beginning heights for every branch.}
  \label{fig:moderate_beg_ht_comparison}
\end{figure}

Figure \ref{fig:moderate_beg_ht_comparison} compares observed beginning heights with our simulations - the reproduction is satisfactory except for asteroidal and iron-rich material at very low velocities ($V_{init} < \SI{12}{\kilo \metre \per \second}$) where only larger meteoroids are visible. These meteoroids are discussed in \cite{jenniskens2016cams} who attribute them to an unexpected population of large and old Poynting-Robertson drag evolved meteoroids at very low semi-major axes ($T_J > 3.2$), indicating collisional lifetimes on the order of \SI{d6}{} years and possibly different physical properties than the rest of the population. Alternatively, the luminous efficiency may change dramatically at lower speeds and our mass model may no longer be valid. We also note that a small number have beginning heights above our modelled range, which may be caused by different physical properties of those meteoroids than modelled, seasonal changes in the atmosphere (see section \ref{subsec:atm_dens_investigation}), or simply observational errors.

The beginning heights of the Perseids and Orioinids match those expected for cometary material, consistent with their cometary origin \citep{borovivcka2005physical}. The Geminids lie between the two discrete branches, suggesting a larger spread in strength/densities and heterogeneity of the meteoroid material, as also suggested by the results of  \cite{borovivcka2009material}, who found the densities of Geminid meteoroids to range between \SI{1000}{\kilogram \per \cubic \metre} and \SI{3000}{\kilogram \per \cubic \metre} for the same mass range. Furthermore, \cite{ceplecha1977five} classified the Geminids as an intermediate type B, between the asteroidal type A and cometary type C, also consistent with our results.

\begin{figure}
  \includegraphics[width=\linewidth]{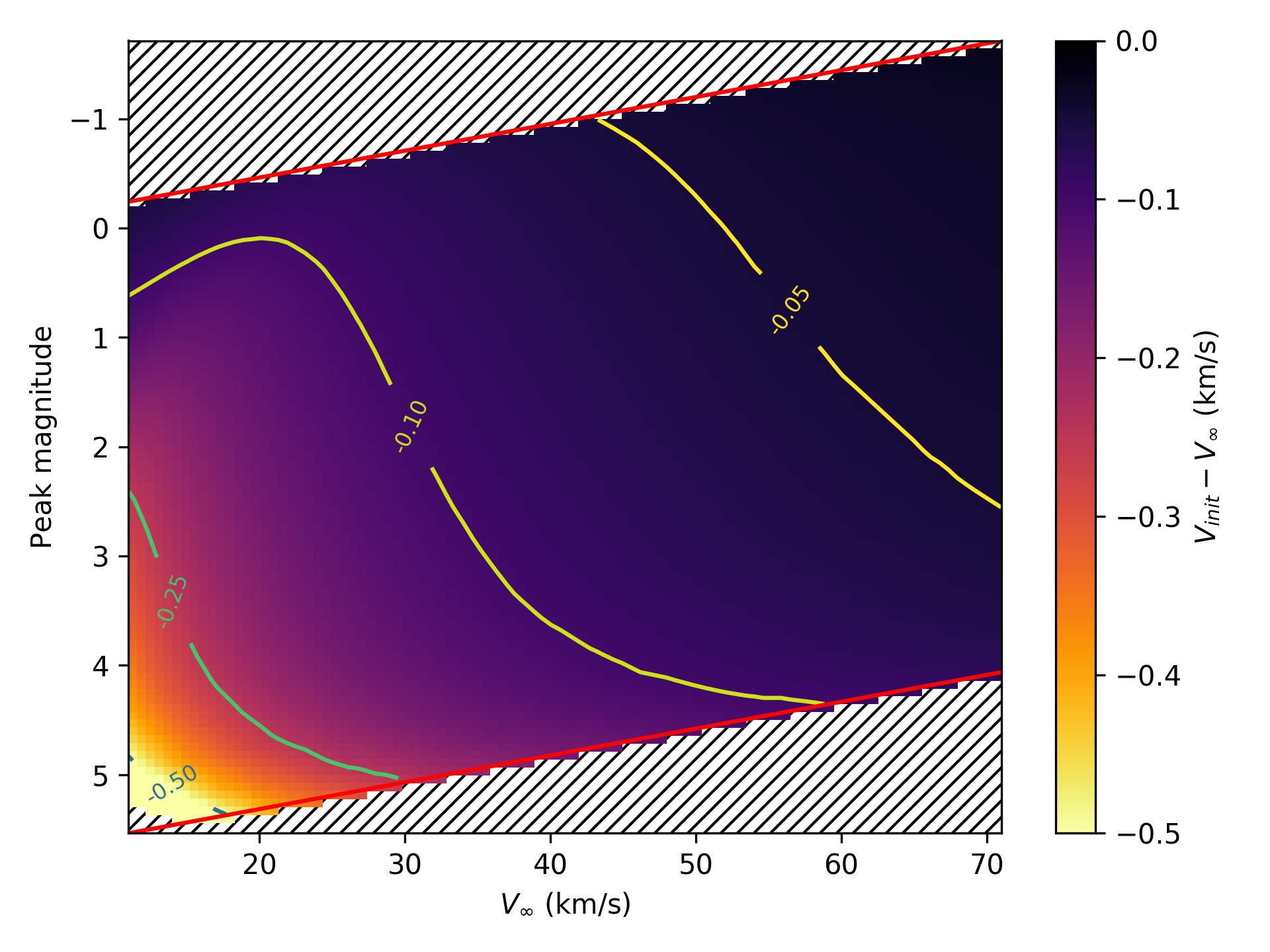}
  \caption{Cometary meteoroids - simulations for the CAMS-type system.}
  \label{fig:sim_moderate_cometary}
\end{figure}

\begin{figure}
  \includegraphics[width=\linewidth]{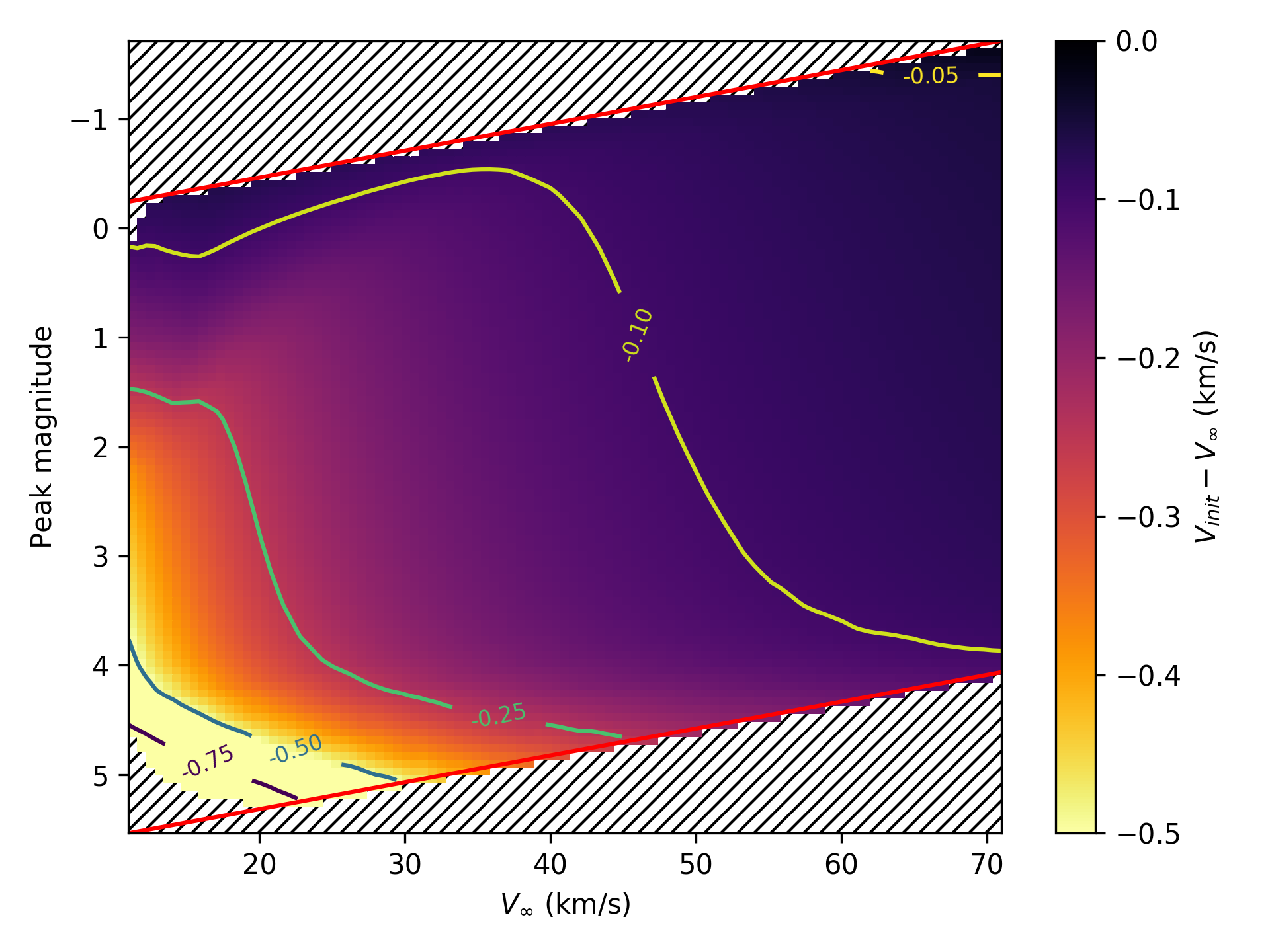}
  \caption{Asteroidal meteoroids - simulations for the CAMS-type system.}
  \label{fig:sim_moderate_asteroidal}
\end{figure}

\begin{figure}
  \includegraphics[width=\linewidth]{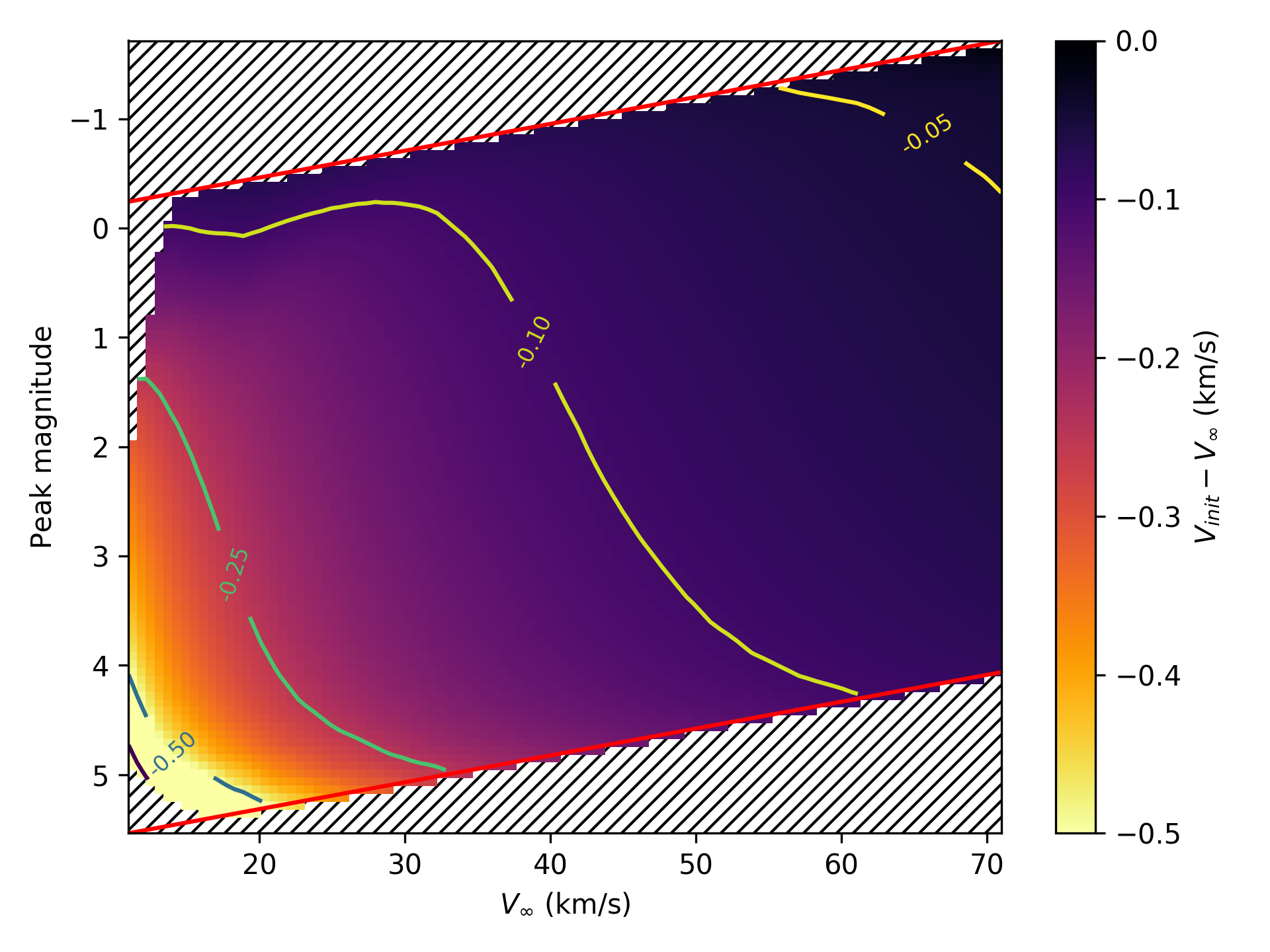}
  \caption{Iron-rich meteoroids - simulations for the CAMS-type system.}
  \label{fig:sim_moderate_iron}
\end{figure}

Figures \ref{fig:sim_moderate_cometary} to \ref{fig:sim_moderate_iron} show the model differences between the initial and pre-atmosphere velocities for various meteoroid types. Compared to the CAMO influx system, the overall differences in velocities are similar, even though the beginning heights are lower, since the meteoroid masses are larger. At high velocities, cometary meteoroids show  velocity differences below \SI{100}{\metre \per \second} down to \SI{20}{\kilo \metre \per \second} when the difference exceeds $\SI{200}{\metre \per \second}$. Asteroidal meteoroids show the highest absolute velocity difference, in excess of \SI{500}{\metre \per \second} for the faintest meteors at low velocities of $v_{\infty} \approx \SI{15}{\kilo \metre \per \second}$. Finally, as with the CAMO influx system, iron-rich meteoroids exhibit velocity differences that are between the other two types.




\subsection{All-sky (SOMN) system}

\begin{figure}
  \includegraphics[width=\linewidth]{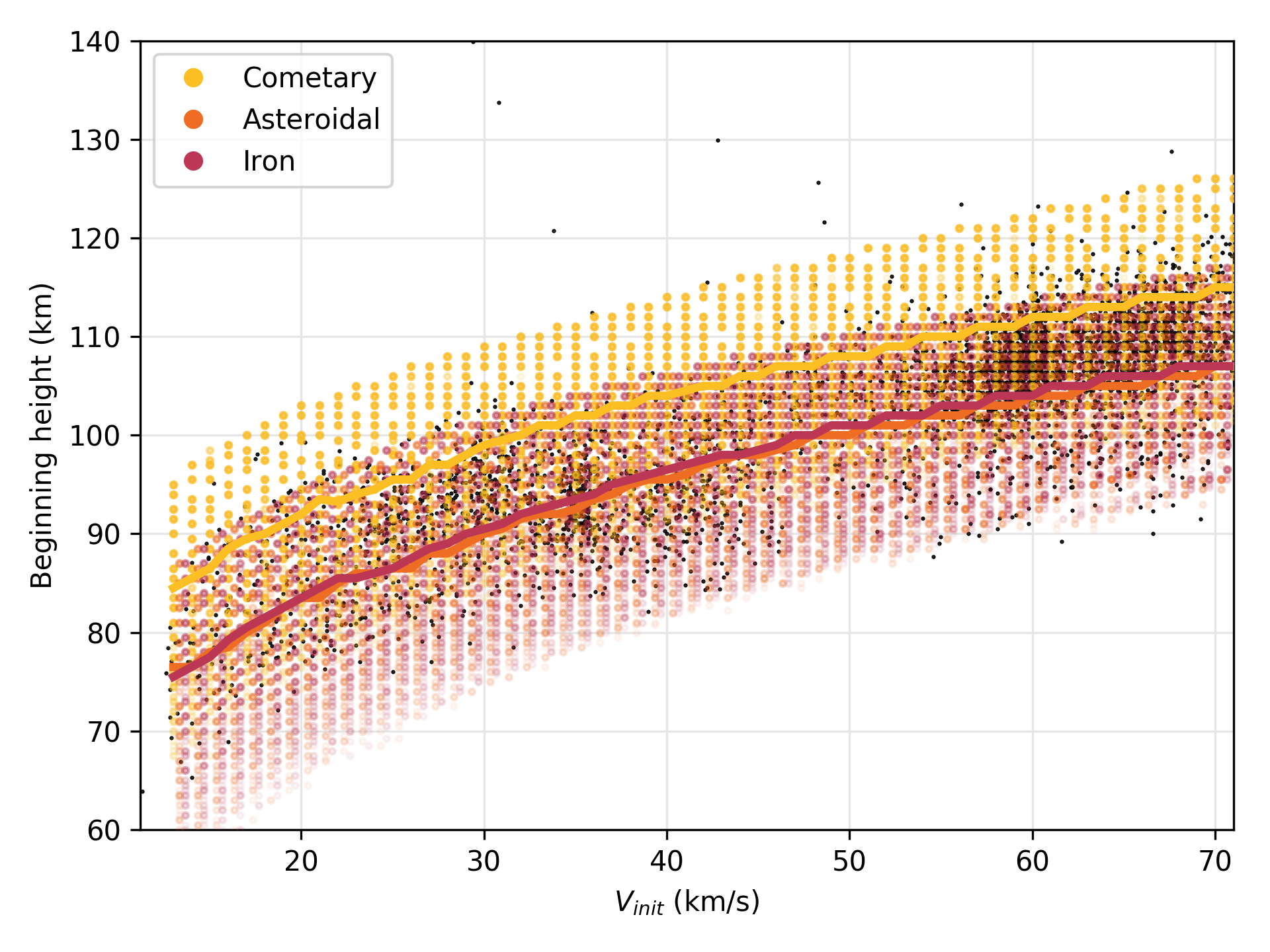}
  \caption{Observed and simulated beginning heights for the all-sky system. Yellow, orange and brown dots represent cometary, asteroidal and iron meteoroids respectively. Thick lines that follow the branches by the middle are median beginning heights for every branch.}
  \label{fig:allsky_beg_ht_comparison}
\end{figure}

Figure \ref{fig:allsky_beg_ht_comparison} shows the comparison of observed beginning heights and our simulations for an optical system with all-sky video sensitivity. The FM model reproduces the trend of beginning heights well for both branches, across all modelled velocities. The only discrepancy is in the upper regions of the cometary branch - simulations indicate that for the assumed physical parameters cometary meteoroids should start higher. This may indicate that the centimetre-size population lacks low-density cometary material, compared to smaller meteoroids seen by more sensitive systems.

\begin{figure}
  \includegraphics[width=\linewidth]{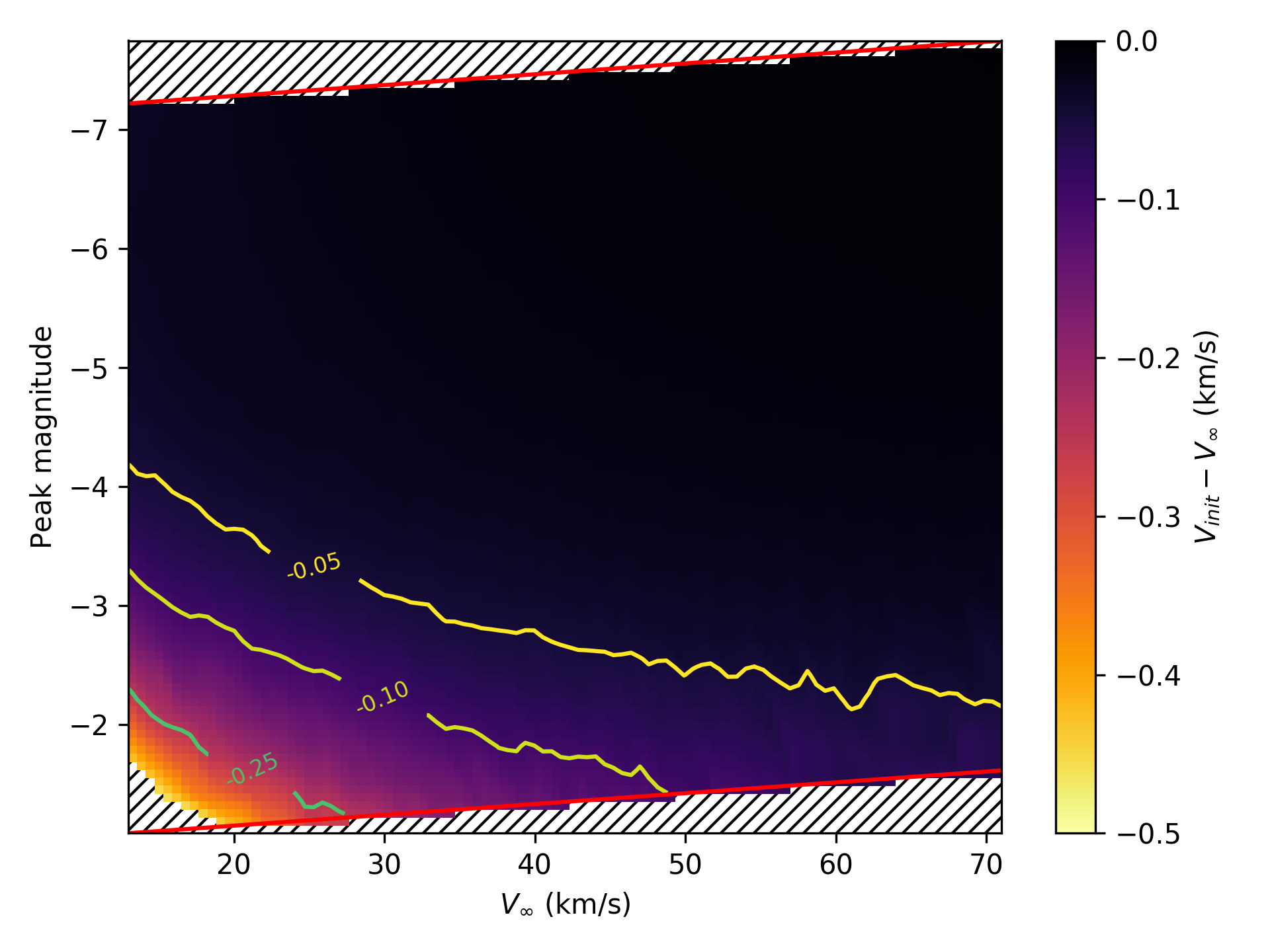}
  \caption{Cometary meteoroids - simulations for all-sky systems.}
  \label{fig:sim_allsky_cometary}
\end{figure}

\begin{figure}
  \includegraphics[width=\linewidth]{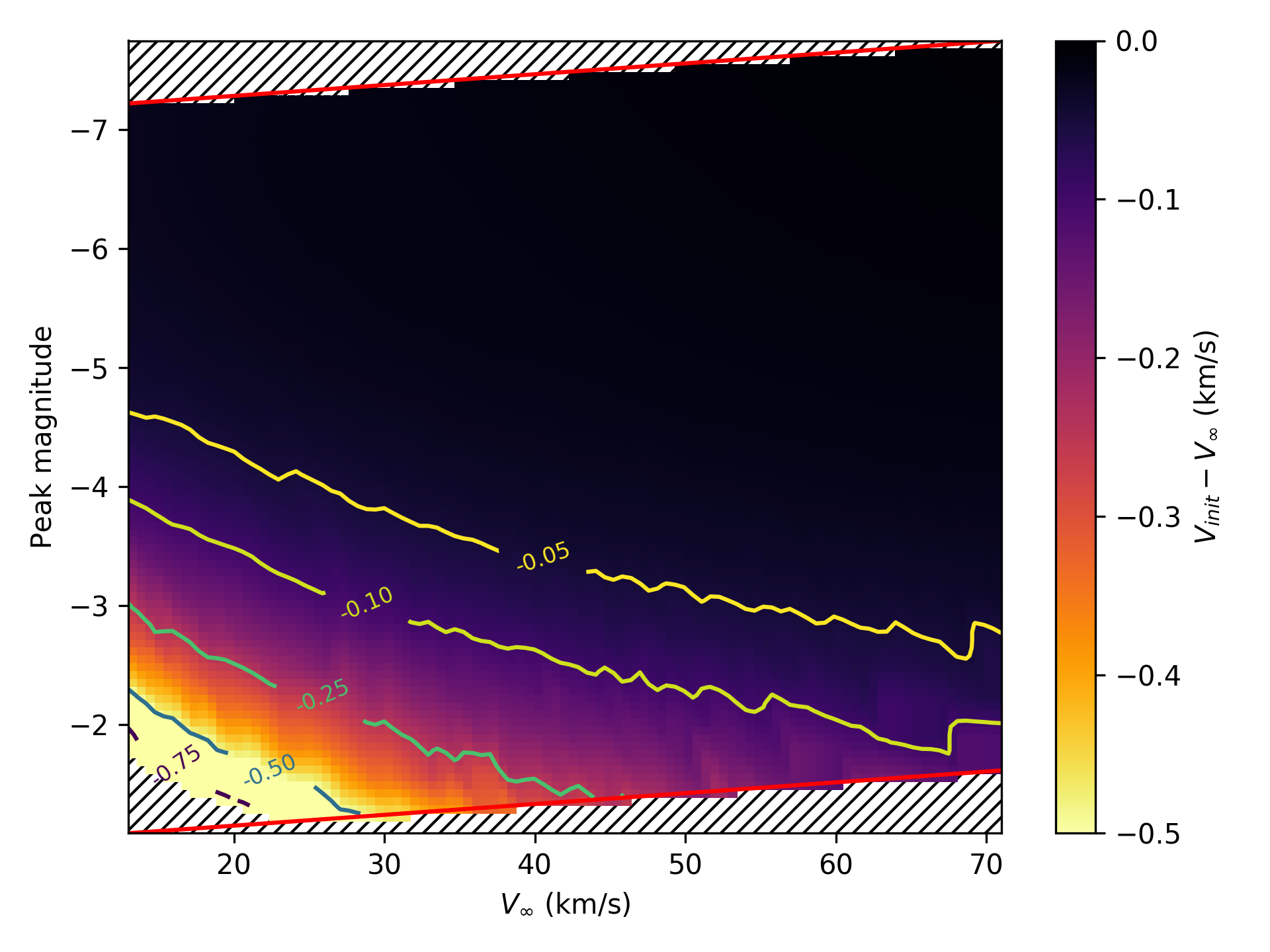}
  \caption{Asteroidal meteoroids - simulations for all-sky systems.}
  \label{fig:sim_allsky_asteroidal}
\end{figure}

\begin{figure}
  \includegraphics[width=\linewidth]{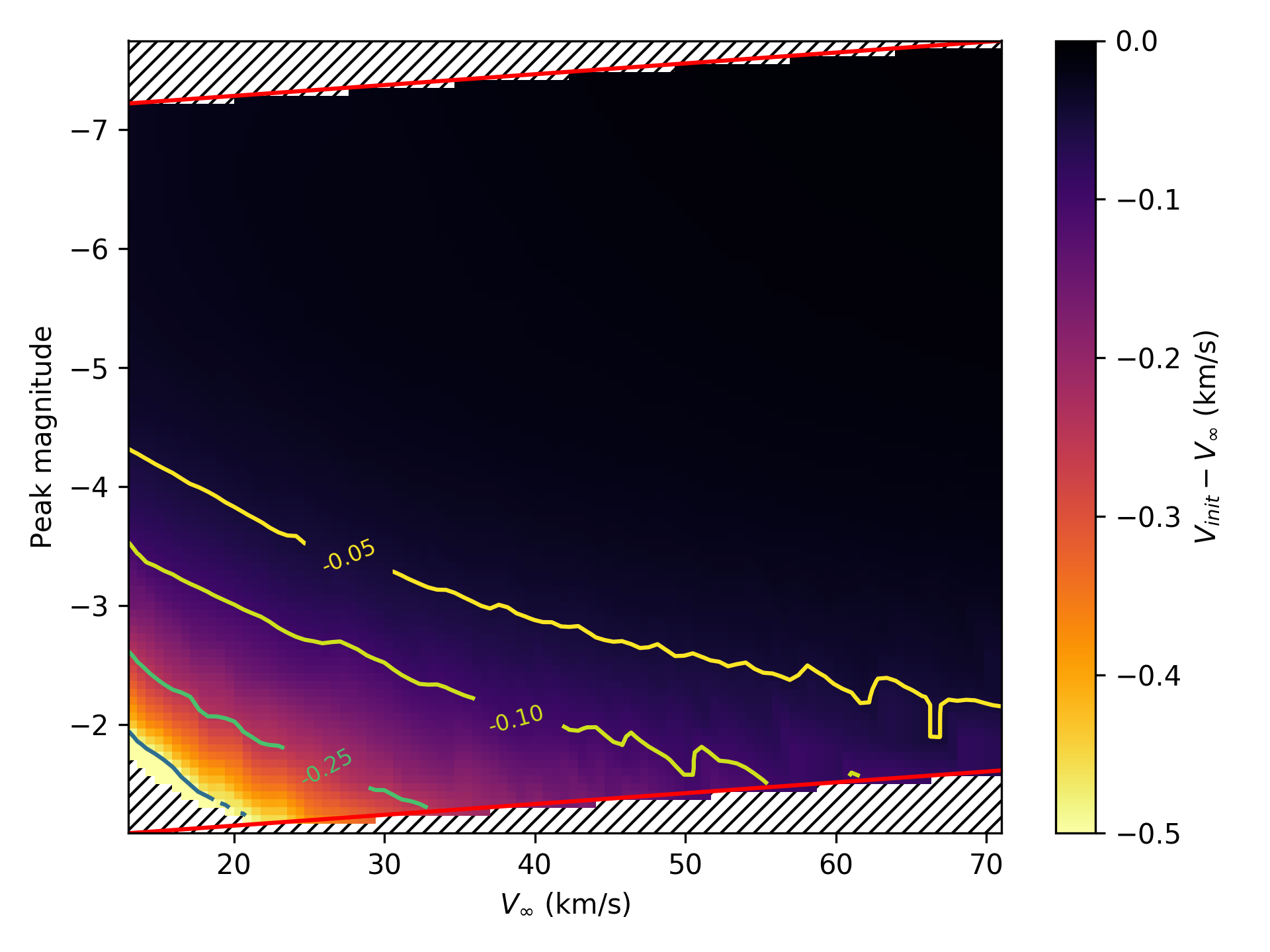}
  \caption{Iron-rich meteoroids - simulations for all-sky systems.}
  \label{fig:sim_allsky_iron}
\end{figure}

Figures \ref{fig:sim_allsky_cometary} through \ref{fig:sim_allsky_iron} show the initial and pre-atmosphere velocity differences. Compared to meteoroids seen by other systems, these have the smallest $\Delta v$, indicating the reduction of the velocity difference with the rise in observed meteoroid masses. For all types of meteoroids with peak magnitudes brighter than $-4^M$, the difference in velocity is below \SI{50}{\metre \per \second}. The difference in velocity is only significant for very low velocity faint meteors, particularly asteroidal meteoroids. It is close to or in excess of $\SI{0.5}{\kilo \metre \per \second}$ for $v_{\infty} < \SI{25}{\kilo \metre \per \second}$ and peak magnitudes below $-2^M$, which are close to the detection limit of the system.




\subsection{Dependence of the velocity difference on the varying atmospheric density} \label{subsec:atm_dens_investigation}

\begin{figure}
  \includegraphics[width=\linewidth]{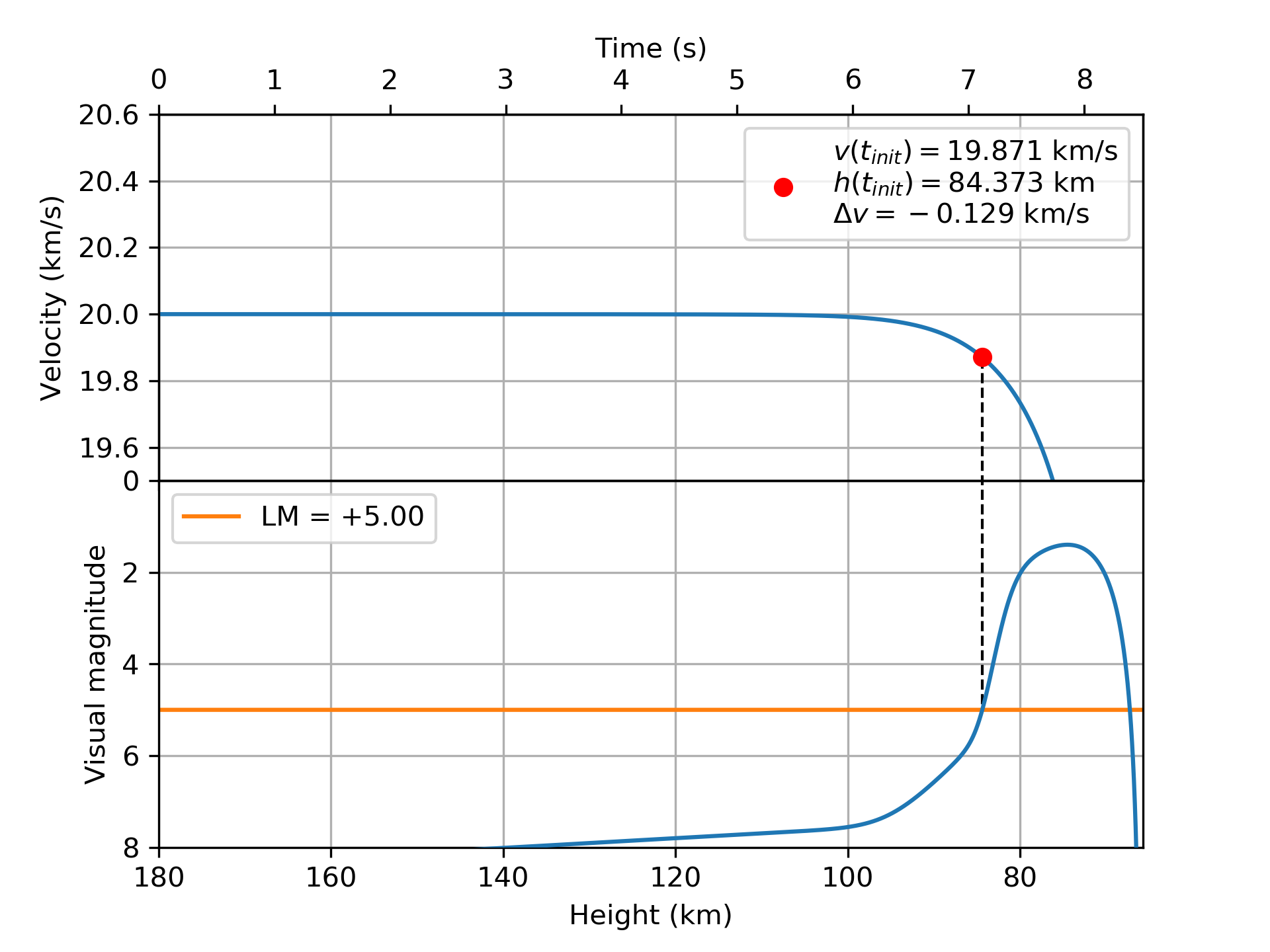}
  \caption{Simulation for 50 \percent lower atmosphere mass density.}
  \label{fig:simulation_vel_mag_50percent_lower}
\end{figure}

\begin{figure}
  \includegraphics[width=\linewidth]{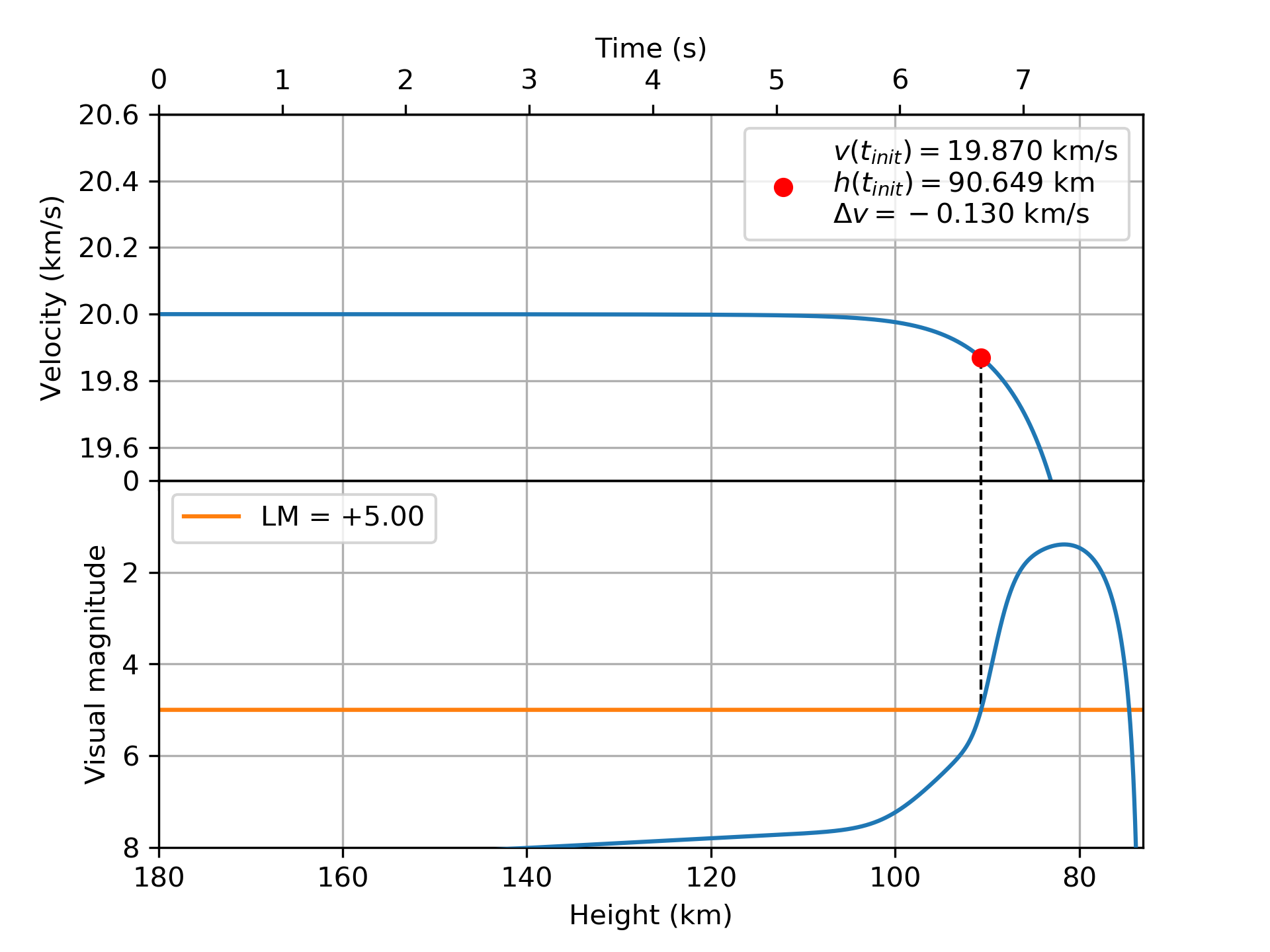}
  \caption{Simulation for 50 \percent higher atmosphere mass density.}
  \label{fig:simulation_vel_mag_50percent_higher}
\end{figure}

Our results may be influenced by latitudinal and seasonal changes in the air mass density at meteor heights, which can vary by up to $50 \percent$ (Dr. Douglas Drob, personal communication). Unfortunately, no currently available models implement these variances in detail. Thus, we investigated the influence of the atmospheric mass density on the velocity difference in two extreme cases, a 50 \percent increase and a 50 \percent decrease in atmospheric mass density. Figures \ref{fig:simulation_vel_mag_50percent_lower} and \ref{fig:simulation_vel_mag_50percent_higher} show simulations of the same meteor as in figure \ref{fig:simulation_vel_mag} but with different values of the atmospheric mass density. Simulations were performed for a $V_{\infty} = \SI{20}{\kilo \metre \per \second}$ cometary meteoroid with a mass of $m = \SI{0.1}{\gram}$, density $\rho_m = \SI{1510}{\kilogram \per \cubic \metre}$ and zenith angle of $Z_G = \ang{45}$, as seen by the simulated CAMS-like system. The results show that the beginning heights shift up or down, but $\Delta v$ remains approximately the same.

\begin{figure}
  \includegraphics[width=\linewidth]{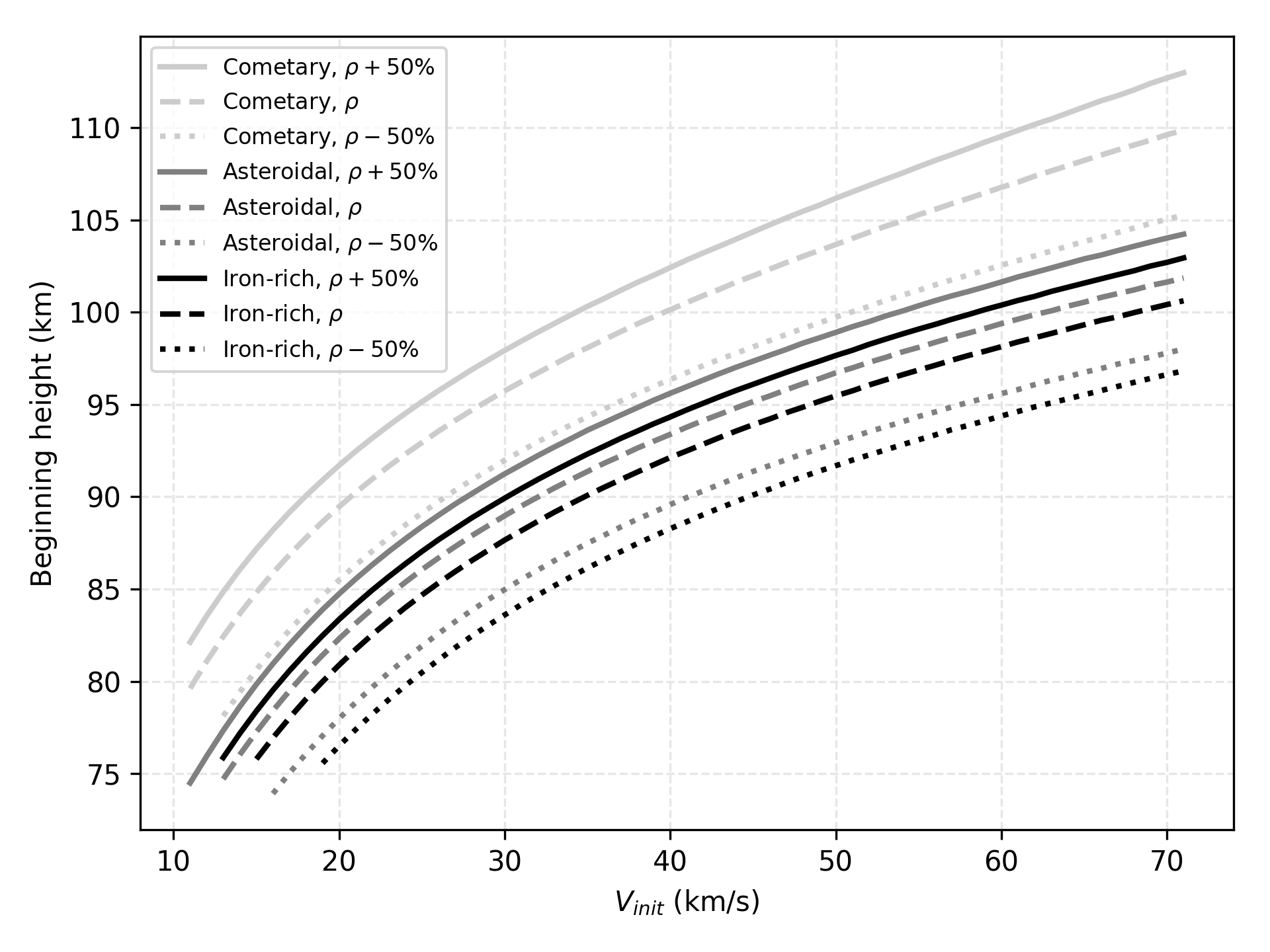}
  \caption{Comparison of beginning heights for 3 meteoroid types and $\pm 50 \percent$ atmosphere mass densities.}
  \label{fig:atm_dens_beg_ht_comparison}
\end{figure}

\begin{figure}
  \includegraphics[width=\linewidth]{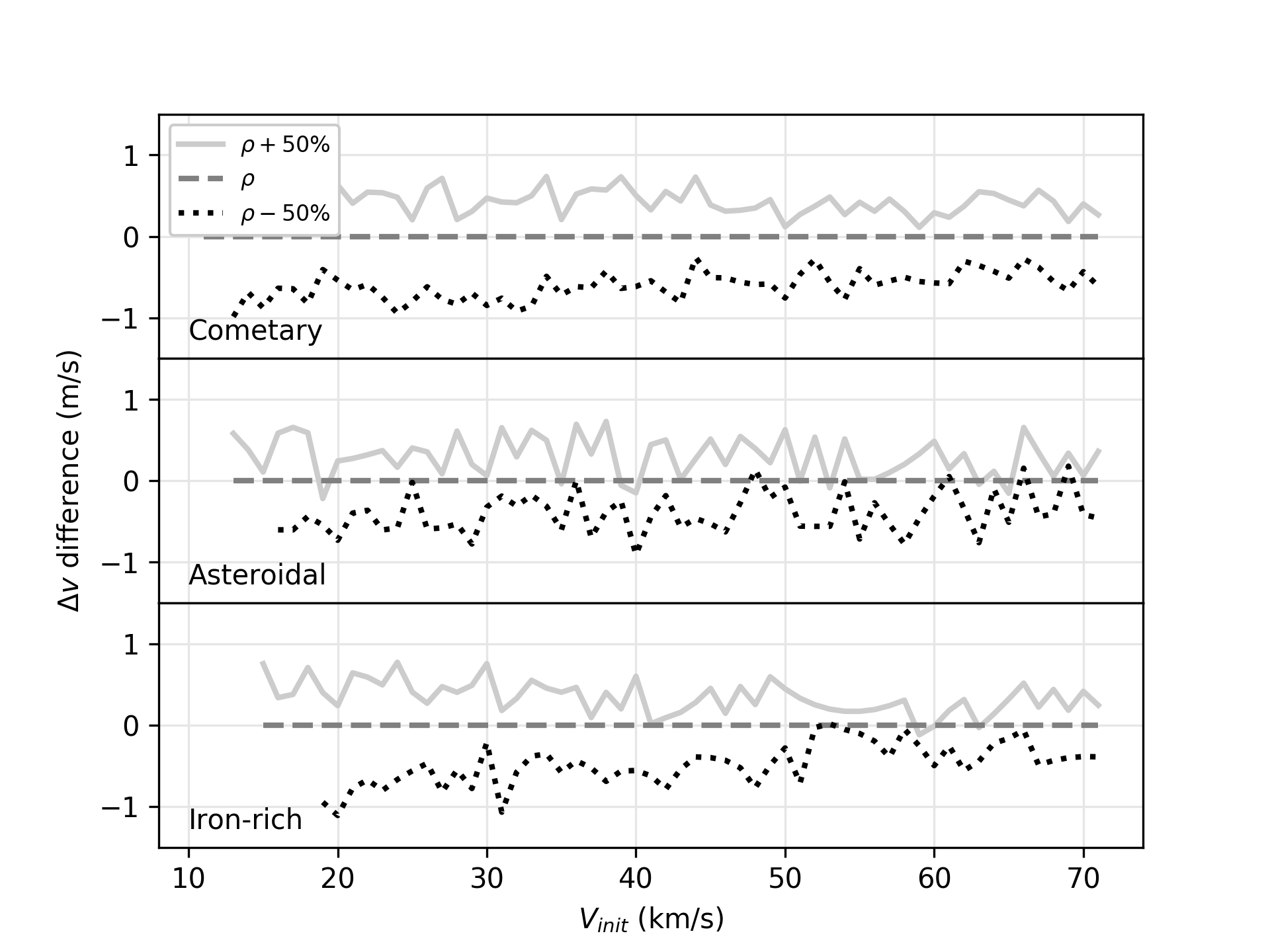}
  \caption{Comparison of differences in $\Delta V$ for 3 meteoroid types and $\pm 50 \percent$ atmosphere mass densities.}
  \label{fig:atm_dens_delta_v_comparison}
\end{figure}

Figure \ref{fig:atm_dens_beg_ht_comparison} shows the comparison of beginning heights for meteoroids of different types. As expected, in the case of a denser atmosphere, meteors start several kilometers higher. Similarly, lower assumed atmospheric mass densities lead to meteors having lower starting heights. In contrast to beginning heights, $\Delta v$ remains virtually unaffected ($< \SI{1}{\metre \per \second}$ difference) by atmosphere density changes of order a factor of two across all velocities and for all meteoroid types, as shown in Figure \ref{fig:atm_dens_delta_v_comparison}.

\section{Model Validation}

As older meteoroid streams are expected to have inherent (physical) dispersions of velocities inside the stream of order several kilometers per second \citep{abedin2017age} it may be difficult to argue that the velocity corrections we are proposing are significant if one considers only the mean velocity of the stream. The largest absolute decelerations before the point of detection are for smaller low-velocity meteoroids which either do not belong to any meteoroid stream or are very dynamically evolved. Due to these unfavourable circumstances, we are only able to validate our model results for the case of the 2011 Draconid outburst. \cite{maslov2011future} and \cite{vaubaillon2011draconid} modelled the ejection of meteoroids from comet 21P/Giacobini-Zinner and predicted that a very young stream of material ejected in 1900 and 1907 will produce an outburst in 2011. Both published predicted a model mean value of meteoroid geocentric velocities of \SI{20.9}{\kilo \metre \per \second} at Earth. 

The outburst was well observed: \cite{toth2012video} observed 62 Draconids from northern Italy, but due to large deceleration they found it difficult to estimate the initial velocity and used a fixed velocity from previous observations by \cite{borovivcka2007atmospheric}. \cite{borovivcka2014spectral} used the \cite{borovivcka2007atmospheric} meteoroid erosion model which takes deceleration into account and matched it to their observations, which enabled them to more accurately estimate pre-atmosphere initial velocities. They obtained a mean geocentric velocity of $V_g = 20.84 \pm 0.15 \SI{}{\kilo \metre \per \second}$, which match the model predictions well. \cite{kero2012mu} used meteor head echo data from the MU radar in Japan and showed directly that their meteoroids decelerate significantly before ablation and detection. After applying a deceleration correction they estimated a mean $V_g = \SI{20.9}{\kilo \metre \per \second}$, also matching the predictions exactly due to the very high velocity precision possible with head echo measurements. 

In contrast, optical observations which did not correct for deceleration before detection estimated geocentric velocities which were $150 - 200 \SI{}{\metre \per \second}$ lower than predicted. \cite{segon2014draconids} determined the initial velocity of 53 video Draconids using the linear deceleration model of \cite{gural2012new} instead of average velocities. The model assumes that the meteoroid starts with an initial velocity of $V_{init}$ and experiences a constant (fixed) deceleration with time. As there is no information about the deceleration before detection, the initial velocity that was measured was the velocity at the beginning height. They found a mean geocentric velocity of $V_g = 20.74 \pm 0.71 \SI{}{\kilo \metre \per \second}$. \cite{trigo20132011} measured the velocity at the beginning of the meteor trail and found $V_g = 20.76 \pm 0.43 \SI{}{\kilo \metre \per \second}$ for 16 manually reduced video Draconids. \cite{jenniskens2016established} used an exponential deceleration model of \cite{whipple1957reduction} with the aim of reconstructing true pre-atmopshere velocities and cite a mean geocentric velocity for the 2011 Draconids of $V_g = \SI{20.7}{\kilo \metre \per \second}$, consistent with other observations measuring only the velocity at the beginning of the visible trail.

The geocentric velocity uncertainty in the three cases above are on the order of hundreds meters per second; however, from our modelling we suggest that they all systematically underestimate the true speeds by $\sim\SI{150}{\metre \per \second}$. Systems used by \cite{segon2014draconids}, \cite{trigo20132011}, and \cite{jenniskens2016established} are comparable to our simulated CAMS-like system, for which the predicted velocity difference for cometary meteoroids at \SI{20}{\kilo \metre \per \second} ranges from $\sim\SI{100}{\metre \per \second}$ to $\sim\SI{500}{\metre \per \second}$, depending on the mass of the meteoroid. This value is a lower boundary as we assume no fragmentation prior to detection, which certainly is not true of the fragile Draconid meteoroids.

\cite{jenniskens1997meteor} noticed the difference in initial velocities of Quadrantids between photographically determined initial velocities (meteor LM $+0^M$, \cite{betlem1997precision} data reduction method) and average velocities of image-intensified video meteors (meteor LM $+6^M$) to be as much as \SI{0.7}{\kilo \metre \per \second}. A large portion of the difference between the two was caused by the overall deceleration of the meteor, but our results suggest that at least \SI{100}{\metre \per \second} of this difference could be due to the inherent observational biases of both systems. Finally, we note that the \SI{200}{\metre \per \second} to \SI{500}{\metre \per \second} initial velocity underestimation for the Geminids described by \cite{hajdukova2017meteoroid} is well explained by our analysis.

\section{Conclusions}

We have modeled the velocities of meteoroids at the top of the atmosphere and compared these to expected measured velocities at the moment of first luminous detection. Our analysis shows that these velocities are expected to differ by a minimum value of order of hundreds of meters per second, the velocity difference being heavily dependent on meteoroid mass, composition, and velocity. In the mass range observed by all-sky fireball networks the difference is almost negligible, while for optical systems detecting typical meteoroid masses smaller than \SI{1}{\gram} the difference is significant and can be in excess of \SI{500}{\metre \per \second}. This implies that increasing the precision of measured initial velocities is not the limiting factor for obtaining high accuracy meteoroid orbits. Improving accuracy requires numerical ablation modelling and additional assumptions about the composition of each meteoroid. As a starting point for such corrections, a table providing empirical lookup corrections per optical system and meteoroid type is given in the appendix.

We have reproduced the observed separation of meteoroids by their beginning heights through ablation modeling and determined that it is largely density dependent, thus allowing classification of meteoroids by their beginning heights into rough density groups, confirming the predictions of \cite{ceplecha1968discrete}. Low-density meteoroids of cometary origin always start at higher altitudes, while asteroidal and iron-rich meteoroids start lower, although the latter two do not differ significantly in their beginning heights. Nevertheless, we notice a discrepancy between our findings and those of \cite{kikwaya2011bulk} for low-density HTC meteors with low beginning heights.\cite{kikwaya2011bulk} found a range of densities, while our model predicts they should all have asteroidal densities. The similarity of beginning heights between the asteroidal and iron-rich group might indicate that they are in fact the same population in terms of bulk density, as proposed by \cite{moorhead2017two}. Notably, that study found that meteoroid densities correlate more strongly with Tisserand parameter than with the \cite{ceplecha1958composition} $K_B$ parameter, which is based on beginning heights. 

Our findings imply a non-negligible systematic observational bias resulting in underestimation of the semi-major axis of low-velocity meteor showers.

\subsection{Note on code availability}

Data files with fit parameters and a Python function which calculates the velocity correction for a given velocity, meteoroid type and system are given on following GitHub web page: \url{https://github.com/dvida/PreatmosphereVelocityCorrection}. Readers are encouraged to contact the authors in the event they are not able to obtain the code on-line.

\section*{Acknowledgements}

Funding for this work was provided through NASA co-operative agreement NNX15AC94A, the Natural Sciences and Engineering Research Council of Canada (Grant no. RGPIN-2016-04433) and the Canada Research Chairs Program. The authors thank Dr. Ji\v{r}i Borovi\v{c}ka for a helpful and detailed review of an earlier version of this manuscript, Prof. Paul Wiegert for fruitful discussions and valuable suggestions, Damir \v{S}egon for clarifications of methods used in their 2011 Draconid paper, and Pete Gural for clarifications of CAMS data format. We also thank Z. Ceplecha, J. Borovi\v{c}ka and P. Spurn{\'{y}} for use of the FM and MILIG codes.




\bibliographystyle{mnras}
\bibliography{mybibfile} 



\appendix

\appendix

\section{Tables of velocity difference fits} \label{appendix:delta_v_fits}

In the supplementary materials we provide a way to compute $\Delta v$ for different systems and meteoroid types using the equation \ref{eq:delta_v_fit}.


\bsp	
\label{lastpage}
\end{document}